\documentclass[aps,superscriptaddress, showpacs,preprintnumbers, superscriptaddress, nofootinbibt,twocolumn]{revtex4}
\usepackage{dcolumn}
\usepackage{bm}
\usepackage{enumerate}
\usepackage{float}
\usepackage{epstopdf}
\usepackage{amsmath}
\usepackage{bm}
\usepackage{amsfonts}
\usepackage{amssymb}
\usepackage{graphicx}
\usepackage{alphalph,mathtools}
\usepackage{etoolbox}

\def\be{\begin{equation}}
\def\ee{\end{equation}}
\def\bea{\begin{eqnarray}}
\def\eea{\end{eqnarray}}

\begin{document}

\title{Energy-dependent noncommutative quantum mechanics}
\author{Tiberiu Harko}
\email{t.harko@ucl.ac.uk}
\affiliation{School of Physics, Sun Yat-Sen University, Guangzhou 510275, People's
Republic of China,}
\affiliation{Department of Physics, Babes-Bolyai University, Kogalniceanu Street,
Cluj-Napoca 400084, Romania,}
\affiliation{Department of Mathematics, University College London, Gower Street, London,
WC1E 6BT, United Kingdom}
\author{Shi-Dong Liang}
\email{stslsd@mail.sysu.edu.cn}
\affiliation{School of Physics, Sun Yat-Sen University, Guangzhou 510275, People's
Republic of China,}
\affiliation{State Key Laboratory of Optoelectronic Material and Technology, and
Guangdong Province Key Laboratory of Display Material and Technology, \\
Sun Yat-Sen University, Guangzhou 510275, People's Republic of China,}
\date{\today }

\begin{abstract}
We propose a model of dynamical noncommutative quantum mechanics in which
the noncommutative strengths, describing the properties of the commutation
relations of the coordinate and momenta, respectively, are arbitrary energy
dependent functions. The Schr\"{o}dinger equation in the energy dependent
noncommutative algebra is derived for a two dimensional system for an
arbitrary potential. The resulting equation reduces in the small energy
limit to the standard quantum mechanical one, while for large energies the
effects of the noncommutativity become important. We investigate in detail
three cases, in which the noncommutative strengths are determined by an
independent energy scale, related to the vacuum quantum fluctuations, by the
particle energy, and by a quantum operator representation, respectively.
Specifically, in our study we assume an arbitrary power laws
energy-dependence of the noncommutative strength parameters, and of their
algebra. In this case, in the quantum operator representation, the Schr\"{o}%
dinger equation can be formulated mathematically as a fractional
differential equation. For all our three models we analyze the quantum
evolution of the free particle, and of the harmonic oscillator,
respectively. The general solutions of the noncommutative Schr\"{o}dinger
equation as well as the expressions of the energy levels are explicitly
obtained.
\end{abstract}

\maketitle
\tableofcontents


\section{Introduction}

It is generally believed today that the description of the space-time as a
manifold $M$, locally modeled as a flat Minkowski space $M_0 =\mathbb{R}%
\times \mathbb{R}^3$, may break down at very short distances of the order of
the Planck length $l_P=\sqrt{G\hbar/c^3}\approx 1.6\times 10^{-33}$ cm,
where $G$, $\hbar$ and $c$ are the gravitational constant, Planck's
constant, and the speed of light, respectively \cite{1}. This assumption is
substantiated by a number of arguments, following from quantum mechanical
and general relativistic considerations, which point towards the
impossibility of an arbitrarily precise location of a physical particle in
terms of points in space-time.

One of the basic principles of quantum mechanics, Heisenberg's uncertainty
principle, requires that a localization $\Delta x$ in space-time can be
reached by a momentum transfer of the order of $\Delta p=\hbar /\Delta x$,
and an energy of the order of $\Delta E=\hbar c/\Delta x$ \cite{1a, Landau, 1b}. On
the other hand, the energy $\Delta E$ must contain a mass $\Delta m$, which,
according to Einstein's general theory of relativity, generates a
gravitational field. If this gravitational field is so strong that it can
completely screen out from observations some regions of space-time, then its
size must be of the order of its Schwarzschild radius $\Delta R\approx
G\Delta m/c^2$. Hence we easily find $\Delta R\approx G\Delta
E/c^4=G\hbar/c^3\Delta x$, giving $\Delta R\Delta x\approx G\hbar /c^3$.
Thus the Planck length appears to give the lower quantum mechanically limit
of the accuracy of position measurements \cite{2}. Therefore the combination
of the Heisenberg uncertainty principle with Einstein's theory of general
relativity leads to the conclusion that at short distances the standard
concept of space and time may lose any operational meaning.

On the other hand, the very existence of the Planck length requires that the
mathematical concepts for high-energy (short distance) physics have to be
modified. This follows from the fact that classical geometrical notions and
concepts may not be well suited for the description of physical phenomena at
very short distances. Moreover, some drastic changes are expected in the
physics near the Planck scale, with one important and intriguing effect
being the emergence of the non-commutative structure of the space-time. The
basic idea behind space-time non-commutativity is very much inspired by
quantum mechanics. A quantum phase space is defined by replacing canonical
position and momentum variables $x^{\mu}$, $p^{\nu}$ with Hermitian
operators that obey the Heisenberg commutation relations,
\begin{equation}
\lbrack \widehat{x}^{\mu },\widehat{p}^{\nu }] =i\hbar \delta ^{\mu \nu }.
\end{equation}

Hence the phase space becomes smeared out, and the notion of a point is
replaced with that of a Planck cell. The generalization of commutation
relations for the canonical operators (coordinate-momentum or
creation-annihilation operators) to non-trivial commutation relations for
the coordinate operators was performed in \cite{3} and \cite{4}, where it
was first suggested that the coordinates $x^{\mu}$ may be noncommutating
operators, with the six commutators being given by
\begin{equation}
\left[\widehat{x}_{\mu},\widehat{x}_{\nu}\right]=i\frac{a^2}{\hbar}L_{\mu
\nu},
\end{equation}
where $a$ is a basic length unit, and $L_{\mu \nu}$ are the generators of
the Lorentz group. In this approach, Lorentz covariance is maintained, but
the translational invariance is lost. A rigorous mathematical approach to
non-commutative geometry was introduced in \cite{5,6,7,8}, by generalizing
the notion of a differential structure to arbitrary $C^*$ algebras, as well
as to quantum groups and matrix pseudo-groups. This approach led to an
operator algebraic description of non-commutative space-times, based
entirely on algebras of functions.

Since at the quantum level non-commutative space-times do appear naturally
when gravitational effects are taken into account, their existence must also
follow from string theory. In \cite{9} it was shown that if open strings
have allowed endpoints on D-branes in a constant $B$-field background, then
the endpoints live on a non-commutative space with the commutation relations
\begin{equation}  \label{e1}
\lbrack \widehat{x}^{\mu },\widehat{x}^{\nu }] =i\theta ^{\mu \nu },
\end{equation}
where $\theta ^{\mu \nu}$ is an antisymmetric \textit{constant} matrix, with
components $c$-numbers with the dimensionality (length)$^{-2}$. More
generally, a similar relation can also be imposed on the particle momenta,
which generates a noncommutative algebra in the momentum space of the form
\begin{equation}
\left[\widehat{p}_{\mu},\widehat{p}_{\nu}\right]=i\eta _{\mu \nu},
\end{equation}
where $\eta _{\mu \nu}$ are constants. In \cite{10a} noncommutative field
theories with commutator of the coordinates of the form $\left[%
x^{\mu},x^{\nu}\right]=i\Lambda ^{\mu \nu}_{\;\;\;\;\omega }x^{\omega}$ have
been studied. By considering $\Lambda$ a Lorentz tensor, explicit Lorentz
invariance is maintained, a free quantum field theory is not affected. On
the other hand, since invariance under translations is broken, the
conservation of energy-momentum tensor is violated, and a new law expressed
by a Poincar\'{e}-invariant equation is obtained. The $\lambda \phi ^4$
quantum field theory was also considered. It turns out that the usual UV
divergent terms are still present in this model. Moreover, new type of terms
also emerge that are IR divergent, violates momentum conservation and lead
to corrections to the dispersion relations.

The physical implications and the mathematical properties of the
non-commutative geometry have been extensively investigated in \cite%
{10,11,12,13,13a,14,14a,15, 15n, 15a, 16,17, 17a, 18,19,20,21,22,23,89,24,25,26, 26a,26b,26c,26d,26e,26f}%
. In the case when $\left [\widehat{p}_i,\widehat p_j\right] = 0$, the
noncommutative quantum mechanics goes into the usual one, described by the
non-relativistic Schr\"{o}dinger equation,
\begin{equation}
H \left(\tilde{x},p\right)\psi \left(\tilde{x}\right)=E\psi \left(\tilde{x}%
\right),
\end{equation}
where $\tilde{x}^{\mu}=x^{\mu}-(1/2)\theta ^{\mu \nu}p_{\nu}$ \cite{27}. In
the presence of a constant magnetic field $B$ and an arbitrary central
potential $V(r)$, with Hamiltonian
\begin{equation}
\widehat{H}=\frac{\widehat{p}^2}{2m}+V(r),
\end{equation}
the operators $\widehat{p}$, $\widehat{x}$ obey the commutation relations
\cite{27}
\begin{eqnarray}
\left[\widehat{x}^1, \widehat{x}^2\right]= i\theta, \;\; \left[\widehat{x}%
^{\mu},\widehat{p}_{\nu}\right]= i \hbar \delta ^{\mu}_{\nu}, \;\; \left[%
\widehat{p}_1,\widehat{p}_2\right] &=& i\frac{e}{c}B.
\end{eqnarray}
Several other types of noncommutativity, extending the canonical one, have
also been proposed. For example, in \cite{28}, a three-dimensional
noncommutative quantum mechanical system with mixing spatial and spin
degrees of freedom was investigated. In this study it was assumed that the
noncommutative spatial coordinates $\widehat{x}^i$, the conjugate momenta $%
\widehat{p}^ i$, and the spin variables $\widehat{s}^i$ obey the nonstandard
Heisenberg algebra

\begin{eqnarray}
\left[\widehat{x}^i,\widehat{x}^j\right]= i\theta ^2\epsilon ^{ijk}\widehat{s%
}^k, \;\; \left[\widehat{x}^i,\widehat{p}_j\right]= i\delta ^i_j, \;\; \left[%
\widehat{p}_i,\widehat{p}_j\right]=0,
\end{eqnarray}
and
\begin{eqnarray}
\left[\widehat{x}^i,\widehat{s}^j\right]= i\theta \epsilon ^{ijk}\widehat{s}%
^k, \;\; \left[\widehat{s}^i,\widehat{s}^j\right]= i\epsilon ^{ijk}\widehat{s%
}^k,
\end{eqnarray}
respectively, where $\theta \in \mathbb{R}$ is the parameter of the
noncommutativity. A classical model of spin noncommutativity was
investigated in \cite{29}. In the nonrelativistic case, the Poisson brackets
between the coordinates are proportional to the spin angular momentum. The
quantization of the model leads to the noncommutativity with mixed spatial
and spin degrees of freedom. A modified Pauli equation, describing a spin
half particle in an external electromagnetic field was also obtained, and it
was shown that in spite of the presence of noncommutativity and nonlocality,
the model is Lorentz invariant. Other physical and mathematical implications
of spin noncommutativity were investigated in \cite{30,31,32}

A model of dynamic position-dependent noncommutativity, involving the
complete algebra of noncommutative coordinates
\begin{equation}
\left[\widehat{x}^{\mu},\widehat{x}^{\nu}\right]=i\omega ^{\mu \nu}\left(%
\widehat{x}\right),
\end{equation}
was proposed in \cite{33}, an further investigated in \cite{34}. In \cite{34}
a system consisting of two interrelated parts was analyzed. The first
describes the physical degrees of freedom with the coordinates $x^1$ and $%
x^2 $, while the second corresponds to the noncommutativity $\eta $, which
has a proper dynamics. It turns out that after quantization, the commutator
of two physical coordinates is proportional to an arbitrary function of $%
\eta $. An interesting feature of this model is the dependence of
nonlocality on the energy of the system, so that the increase of the energy
leads to the increase in nonlocality. The physical properties of systems
with dynamic noncommutativity were considered in \cite{35,36,37, 37a, 37b,
38, 39,40}.

A quantum mechanical system on a noncommutative space for which the
structure constant is explicitly time-dependent was investigated in \cite%
{40a}, in a two-dimensional space with nonvanishing commutators for the
coordinates $X$, $Y$ and momenta $P_x$, $P_y$ given by
\begin{subequations}
\begin{eqnarray}
[X,Y ] &= &i\theta (t), \left[P_x,P_y\right] = i\Omega (t), \\
\left[X,P_x\right] &=& \left[Y,P_y\right] = i\hbar + i \frac{\theta(t)\Omega
(t)}{ 4\hbar}.
\end{eqnarray}
Any autonomous Hamiltonian on such a space acquires a time-dependent form in
terms of the conventional canonical variables. A generalized version of
Heisenberg's uncertainty relations for which the lower bound becomes a
time-dependent function of the background fields was also obtained. For a
two dimensional harmonic oscillator, after performing the Bopp shift, the
Hamiltonian becomes times dependent, and is given by \cite{40a}

\end{subequations}
\begin{eqnarray}
H(t) &=&\frac{1}{2m_{e}(t)}\left(p^2_x + p^2_y\right)  \notag \\
&&+\frac{k_{e}(t)}{2}\left(x^2 +y^2\right)+ B_{e}(t) L_{z},
\end{eqnarray}
where
\begin{subequations}
\begin{eqnarray}
\hspace{-0.8cm}\frac{1}{m_{e}(t)} &=&\frac{1}{m}+\frac{m\omega ^2}{4 \hbar
^{2}} \theta ^2(t), k_{e}(t) = m\omega ^2 +\frac{\Omega^2(t)}{4 m \hbar ^{2}}%
, \\
\hspace{-0.8cm}B_{e} &=& \frac{m\omega ^2\theta (t)}{2 \hbar} +\frac{\Omega
(t)}{2\hbar m}, L_{z}= \left(p_x y - xp_y\right).
\end{eqnarray}

From a general physical point of view we can interpret the noncommutativity
parameters $\theta ^{\mu \nu}$ and $\eta _{\mu \nu}$ as describing \textit{%
the strength} of the noncommutative effects exerted in an interaction. In
this sense they are the analogues of the \textit{coupling constants} in
standard quantum field theory.

It is a fundamental assumption in quantum field theory that the properties
of a physical system (including the underlying force laws) change when
viewed at \textit{different distance scales}, and these changes are \textit{%
energy dependent}. This is the fundamental idea of the \textit{%
renormalization group method}, which has found fundamental application in
quantum field theory, elementary particle physics, condensed matter etc.
\cite{41}.

It is the main goal of the present paper \textit{to introduce and analyze a
dynamic noncommutative model of quantum mechanics, in which the
noncommutative strengths $\theta ^{\mu \nu}$ and $\eta _{\mu \nu}$ are
energy-dependent quantities}. This would imply \textit{the existence of
several noncommutative scales} that range from the energy level of the
standard model, where the low energy scales of the physical systems reduce
the general noncommutative algebra to the standard Heisenberg algebra, and
ordinary quantum mechanics, to the Planck energy scale. On energy scales of
the order of the Planck energy $E_P=\sqrt{\hbar c^5/G}\approx 1.22\times
10^{19}$ GeV, the noncommutative effects become maximal. Under the
assumption of the energy dependence of the noncommutative parameters, with
the help of the generalized Seiberg-Witten map, we obtain the general form
of the Schr\"{o}dinger equation describing the quantum evolution in an
energy dependent geometry. he noncommutative effects can be included in the
equation via a generalized quantum potential, which contains an effective
(analogue) magnetic field, as well as an effective elastic constant, whose
functional forms are determined by the energy dependent noncommutative
strengths.

The possibility of an energy-dependent Schr\"{o}dinger equation was first
suggested by Pauli \cite{End1}, and it was further considered and
investigated extensively (see \cite{End2,End3, End3a, End4, End5, End6} and
references therein). Generally, the nonlinearity induced by the energy
dependence requires modifications of the standard rules of quantum mechanics
\cite{End2}. In the case of a linear energy dependence of the potential for
confining potentials the saturation of the spectrum is observed, which
implies that with the increase of the quantum numbers the eigenvalues reach
an upper limit \cite{End4}. The energy-dependent Schr\"{o}dinger equation
was applied to the description of heavy quark systems in \cite{End3}, where
for a linear energy dependence the harmonic oscillator was studied as an
example of a system admitting analytical solutions. A new quark interaction
was derived in \cite{End3a}, by means of a Tamm-Dancoff reduction, from an
effective field theory constituent quark model. the obtained interaction is
nonlocal and energy dependent. Moreover, it becomes positive and rises up to
a maximum value when the interquark distance increases. The quantum
mechanical formalism for systems featuring energy-dependent potentials was
extended to systems described by generalized Schr\"{o}dinger equations that
include a position-dependent mass in \cite{End6}. Modifications of the
probability density and of the probability current need the adjustments in
the scalar product and the norm. The obtained results have been applied to
the energy-dependent modifications of the Mathews-Lakshmanan oscillator, and
to the generalized Swanson system.

From a physical point of view we can assume that the energy dependent
noncommutative effects can be described by two distinct energy scales. One
is the energy scale of the spacetime quantum fluctuations, generated by the
vacuum background and the zero point energy of the quantum fields. The
noncommutativity is then essentially determined by this energy scale, which
is independent of the particle energy. This is the first explicit model we
are considering, a two energy scales model, in which the energy of the
quantum fluctuations and the particle energy evolve in different and
independent ways. The alternative possibility, in which the noncommutative
strengths are dependent on the particle energy only, is also investigated.
We consider the quantum evolution of the free particle and of the harmonic
oscillator in these cases, and the resulting energy spectrum and wave
functions are determined. The particle oscillation frequencies are either
dependent on the vacuum fluctuation energy scale, or they have an explicit
dependence on the particle energy. In the limiting case of small energies we
recover the standard results of quantum mechanics.

As a simple application of the developed general formalism we consider the
case in which the noncommutative strengths $\eta $ and $\theta $ are power
law functions of energy, with arbitrary exponents. However,\textit{the
quantization of such systems}, in which we associate an operator to the
energy, \textit{requires the mathematical/physical interpretation of
operators of the form $\partial ^{\alpha}/\partial t^{\alpha}$, where $%
\alpha $ can have arbitrary real) values, like, for example, $\alpha =1/2$, $%
\alpha =5/4$ etc.} These types of problems belong to the field of \textit{%
fractional calculus} \cite{fract1,fract2,Hermann}, whose physical
applications have been intensively investigated. In particular, the
mathematical and physical properties of the \textit{fractional Schr\"{o}%
dinger equation}, whose introduction was based on a purely phenomenological
or abstract mathematical approach, have been considered in detail in \cite%
{fSch1,fSch2,fSch3,fSch4,fSch5,fSch6,fSch7,fSch8,fSch9,fSch9a,
fSch10,fSch11,fn1,fn2,fn3,fn4,fn5,fn6,fn7,fn8}. It is interesting to note
that the present approach gives a physical foundation for the mathematical
use of fractional derivatives in quantum mechanics, as resulting from the
noncommutative and energy dependent structure of the spacetime. We present
in full detail the fractional Schr\"{o}dinger equations obtained by using
two distinct quantization of the energy (the time operator and the
Hamiltonian operator approach, respectively), and we investigate the quantum
evolution of the free particle and of the harmonic oscillator in the time
operator formalism for a particular simple choice of the energy dependent
noncommutativity strength parameters.

The present paper is organized as follows. We introduce the energy-dependent
non-commutative quantum geometry, its corresponding algebra, and the
Seiberg-Witten map that allows to construct the noncommutative set of
variables from the commutative ones in Section~\ref{sect1}. The Schr\"{o}%
dinger equation describing the quantum evolution in the energy-dependent
noncommutative geometry is obtained in Section~\ref{sect2}, where the form
of the effective potential induced by the noncommutative effects is also
obtained. Three relevant physical and mathematical mechanisms that could
induce energy-dependent quantum behaviors in noncommutative geometry are
discussed in Section~\ref{sect3}, and their properties are explored in the
framework of a particular model in which the noncommutativity parameters
have a power law dependence on energy. The quantum dynamics of a free
particle and of the harmonic oscillator in the Spacetime Quantum
Fluctuations model is analyzed in Section~\ref{sect4}, while the same
physical systems are analyzed in the Energy Coupling model in Section~\ref%
{sect5}. The fractional Schr\"{o}dinger equations for the quantum evolution
of general quantum systems in the Energy Operator approach are presented in
Section~\ref{sect6}, where the dynamics of the free particles and of the
harmonic oscillator are analyzed in detail. A brief review of the fractional
calculus is also presented. We discuss and conclude our results in Section~%
\ref{sect7}.

\section{Energy-dependent noncommutative geometry and algebra}\label{sect1}

In the present Section we will introduce the basic definitions, conventions
and relations for an energy-dependent generalization of the noncommutative
geometry and algebra of physical variables, valid in the high energy/small
distances regime. In high energy physics theoretical models where both the
coordinate and momentum space noncommutativity is taken into account, in a
four-dimensional space the coordinates and momenta satisfy the following
algebra,

\end{subequations}
\begin{subequations}
\begin{eqnarray}
\lbrack \widehat{x}^{\mu },\widehat{x}^{\nu }] &=&i\theta ^{\mu \nu },
\label{16_1} \\
\lbrack \widehat{p}^{\mu },\widehat{p}^{\nu }] &=&i\eta ^{\mu \nu },
\label{16_2} \\
\lbrack \widehat{x}^{\mu },\widehat{p}^{\nu }] &=&i\hbar \Delta ^{\mu \nu }~,
\label{16_3}
\end{eqnarray}%
where the \textit{noncommutativity strength parameters $\theta ^{\mu \nu }$
and $\eta ^{\mu \nu }$ are antisymmetric}. Due to the commutation relation
given by Eq.~(\ref{16_3}), this algebra is consistent with ordinary Quantum
Mechanics. We shall assume in the following that both matrices $\theta ^{\mu
\nu}$ and $\eta ^{\mu \nu}$ are invertible, and moreover the matrix $\Sigma
^{\mu \nu}=\delta ^{\mu \nu}+\left(1/\hbar ^2\right)\theta ^{\mu \alpha}\eta
^{\nu}_{\alpha}$ is also invertible \cite{17a}. Under a linear
transformation of the form, also called the $D$ map,
\begin{equation}  \label{lintrans}
\widehat{x}^{\mu}=A^{\mu}_{\nu}x^{\nu}+B^{\mu}_{\nu}p^{\nu},\widehat{p}%
^{\mu}=C^{\mu}_{\nu}x^{\nu}+D^{\mu}_{\nu}p^{\nu},
\end{equation}
where $\mathbf{\mathrm{A}}$, $\mathbf{\mathrm{B}}$, $\mathbf{\mathrm{C}}$,
and $\mathbf{\mathrm{D}}$ are real constant matrices, the noncommutative
algebra (\ref{16_1})-(\ref{16_3}) can be mapped to the usual Heisenberg
algebra, $\left[x^{\mu},x^{\nu}\right]=0$, $\left[p^{\mu},p^{\nu}\right]=0$,
and $\left[x^{\mu},p^{\nu}\right]=i\hbar \delta ^{\mu \nu}$, respectively
\cite{17a}. The matrices $\mathbf{\mathrm{A}}$, $\mathbf{\mathrm{B}}$, $%
\mathbf{\mathrm{C}}$, and $\mathbf{\mathrm{D}}$ satisfy the equations $%
\mathbf{\mathrm{A}}\mathbf{\mathrm{D}}^T-\mathbf{\mathrm{BC}}^T=\mathbf{%
\mathrm{I}}_{d\times d}$, $\mathbf{\mathrm{A}}\mathbf{\mathrm{B}}^T-\mathbf{%
\mathrm{B}}\mathbf{\mathrm{A}}^T=\Theta/\hbar$, and $\mathbf{\mathrm{C}}%
\mathbf{\mathrm{D}}^T-\mathbf{\mathrm{D}}\mathbf{\mathrm{C}}^T=\mathbf{%
\mathrm{N}}/\hbar$ \cite{17a}, where $\Theta $ and $N$ are matrices with the
entries $\theta ^{\mu \nu}$ and $\eta ^{\mu \nu}$, respectively \cite{17a}.
Due to the linear transformations (\ref{lintrans}), the noncommutative
algebra (\ref{16_1})-(\ref{16_3}) admits a Hilbert space representation of
ordinary quantum mechanics. However, it is important to note that the $D$
map is not unique.

In the present paper we generalize the algebra given by Eqs.~(\ref{16_1})-(%
\ref{16_3}) to the case when \textit{the parameters $\theta ^{\mu \nu }$ and
$\eta ^{\mu \nu }$ are energy-dependent functions}, so that

\end{subequations}
\begin{equation}
\theta ^{\mu \nu }=\theta ^{\mu \nu }(E),
\end{equation}
and
\begin{equation}
\eta ^{\mu \nu }=\eta ^{\mu \nu }(E),
\end{equation}
respectively, where $E$ is a general energy parameter whose physical
interpretation depends on the concrete physical problem under consideration.

In the following we will not consider time-like noncommutative relations,
that is, we take $\theta ^{0i}=0$ and $\eta ^{0i}=0$, since otherwise the
corresponding quantum field theory is not unitary.

The parameters $\theta $ and $\eta $ from Eqs. (\ref{16_1})-(\ref{16_3}),
can be represented generally as
\begin{subequations}
\begin{eqnarray}  \label{SED3}
\left(\theta^{ij}\right)&=&\left(
\begin{array}{cc}
0 & \theta(E) \\
-\theta(E) & 0%
\end{array}%
\right), \\
\left(\eta^{ij}\right)&=&\left(
\begin{array}{cc}
0 & \eta(E) \\
-\eta(E) & 0%
\end{array}%
\right), \\
\left(\Delta^{ij}\right)&=&\left(
\begin{array}{cc}
\gamma^{2}+\frac{\theta(E)\eta(E)}{2\gamma^{2}\hbar^{2}} & \frac{%
\theta(E)\eta(E)}{4\gamma^{2}\hbar^{2}} \\
\frac{\theta(E)\eta(E)}{4\gamma^{2}\hbar^{2}} & \gamma^{2}+\frac{%
\theta(E)\eta(E)}{2\gamma^{2}\hbar^{2}}%
\end{array}%
\right),
\end{eqnarray}
where $i,j$ correspond to $x,y$ and $z$. It can be seen that $\theta$ and $%
\eta$ are antisymmetric, but $\Delta$ is symmetric. They are assumed to be
\textit{energy-dependent}, and we take them as independent of the space-time
coordinates. We may set $\gamma=1$ for convenience without losing the basic
physics in the following Section.

Moreover, we also limit our analysis to the $x-y$ plane, where the \textit{%
two-dimensional noncommutative energy-dependent algebra} can be formulated as

\end{subequations}
\begin{subequations}
\begin{eqnarray}
\hspace{-0.5cm}\left[ \widehat{x},\widehat{y}\right] &=&i\theta (E),\left[
\widehat{p}_{x},\widehat{p}_{y}\right] =i\eta (E),\left[ \widehat{x}_{i},%
\widehat{p}_{j}\right] =i\hbar \delta _{ij},  \label{noncommutation_3} \\
\hspace{-0.5cm}&&\left[ E,\widehat{x}_{i}\right] =0,\left[ E,\widehat{p}_{i}%
\right] =0,i=1,2.  \label{nc4}
\end{eqnarray}%
where, in the last three commutation relations, we have denoted $\widehat{x}%
_{1}\equiv \widehat{x}$, $\widehat{x}_{2}\equiv \widehat{y}$, $\widehat{p}%
_{1}\equiv \widehat{p}_{x}$ and $\widehat{p}_{2}\equiv \widehat{p}_{y}$.

Starting from the canonical quantum mechanical Heisenberg commutation
relations 
one can easily verify that the commutation relations Eq. (\ref%
{noncommutation_3}) can be obtained through the linear transformations \cite%
{9,15}
\end{subequations}
\begin{subequations}
\begin{equation}
\left(
\begin{array}{c}
\widehat{x} \\
\widehat{y}%
\end{array}%
\right) =\left(
\begin{array}{c}
x-{\frac{\theta (E)}{\hbar }}p_{y} \\
y%
\end{array}%
\right) ,  \label{linear_1}
\end{equation}
\begin{equation}
\left(
\begin{array}{c}
\widehat{p}_{x} \\
\widehat{p}_{y}%
\end{array}%
\right) =\left(
\begin{array}{c}
p_{x} \\
p_{y}-{\frac{\eta (E)}{\hbar }}x%
\end{array}%
\right) ,  \label{linear_1a}
\end{equation}%
or, equivalently, through the alternative set of linear transformations

\end{subequations}
\begin{subequations}
\begin{equation}
\left(
\begin{array}{c}
\widehat{x} \\
\widehat{y}%
\end{array}%
\right) =\left(
\begin{array}{c}
x \\
y+{\frac{\theta (E)}{\hbar }}p_{x}%
\end{array}%
\right) ,  \label{linear_2}
\end{equation}
\begin{equation}
\left(
\begin{array}{c}
\widehat{p}_{x} \\
\widehat{p}_{y}%
\end{array}%
\right) =\left(
\begin{array}{c}
p_{x}+{\frac{\eta (E)}{\hbar }}y \\
p_{y}%
\end{array}%
\right) .  \label{linear_2a}
\end{equation}
These two types of linear transformations can be combined into a single one,
which simultaneously modifies all coordinates and momenta, and not just $x$
and $p_{y}$ or $y$ and $p_{x}$, as given in Eqs. (\ref{linear_1})-(\ref%
{linear_1a}) and (\ref{linear_2})-(\ref{linear_2a}), respectively.

One possible way of implementing the algebra defined by Eqs. (\ref%
{noncommutation_3}) and (\ref{nc4}) is to construct the noncommutative set
of variables $\left(\widehat{x},\widehat{y},\widehat{p}_{x},\widehat{p}%
_{y}\right)$ from the commutative variables $\left(x,y,p_{x},p_{y}\right)$
by means of linear transformations. This can be generally done by using
\textit{the Seiberg-Witten map}, given by \cite{9,15}

\end{subequations}
\begin{subequations}
\begin{eqnarray}
\left(
\begin{array}{c}
\widehat{x} \\
\widehat{y}%
\end{array}
\right) &=& \left(
\begin{array}{c}
x-{\frac{\theta (E)}{2\hbar }}p_{y} \\
y+{\frac{\theta (E)}{2\hbar }}p_{x}%
\end{array}
\right),  \label{linear_3} \\
\left(
\begin{array}{c}
\widehat{p}_{x} \\
\widehat{p}_{y}%
\end{array}
\right) &=& \left(
\begin{array}{c}
p_{x}+{\frac{\eta (E)}{2\hbar }}y \\
p_{y}-{\frac{\eta (E)}{2\hbar }}x%
\end{array}
\right),  \label{linear_3a}
\end{eqnarray}%
where the canonical variables $\left(x,y,p_{x},p_{y}\right)$ satisfy
Heisenberg commutation relations \cite{9,15},

\end{subequations}
\begin{eqnarray}
\lbrack x,y] &=& [p_{x},p_{y}]=0, \\
\lbrack x_{i},p_{j}] &=& i\hbar \delta _{ij}, i=1,2,
\end{eqnarray}

With the help of transformations (\ref{linear_3}) and (\ref{linear_3a}) we
can immediately recover the two first commutation relations in Eq. (\ref%
{noncommutation_3}). However, the last one takes the form
\begin{equation}
\lbrack \widehat{x}_{i},\widehat{p}_{j}]=i\hbar \Bigg[1+{\frac{\theta
(E)\eta (E)}{4\hbar ^{2}}}\Bigg]\delta _{ij},i=1,2~.
\label{noncommutation_4}
\end{equation}

Comparing Eqs.~(\ref{noncommutation_3}) and (\ref{noncommutation_4}), we
find that the linear transformations given by Eqs.~(\ref{linear_3}) and (\ref%
{linear_3a}) generate an \emph{effective energy dependent Planck constant},
which is a function of the noncommutativity parameters $\theta (E)$ and $%
\eta (E)$, and it is given by \cite{15}
\begin{equation}
\mathit{\hbar }_{eff}=\hbar \left[ 1+\zeta (E)\right] ~,
\label{Planck_constant_1}
\end{equation}%
where $\zeta \equiv \theta (E)\eta (E)/4\hbar ^{2}$. This approach is
consistent with the usual commutative space-time quantum mechanics if we
impose the condition $\xi \ll 1$, expected to be generally satisfied, since
the small noncommutative parameters $\theta $ and $\eta $, $\zeta $ are of
second order.

For the sake of completeness we also present the general case. In the
four-dimensional space-time Eqs. (\ref{linear_3}) and (\ref{linear_3a}) can
be written as \cite{15}
\begin{equation}  \label{linear_4}
\left(
\begin{array}{c}
\widehat{x}^{\mu } \\
\widehat{p}^{\mu }%
\end{array}%
\right) =\left(
\begin{array}{c}
x^{\mu }-{\frac{\theta _{\ \nu }^{\mu }(E)}{2\hbar }}p^{\nu } \\
p^{\mu }+{\frac{\eta _{\ \nu }^{\mu }(E)}{2\hbar }}x^{\nu }%
\end{array}%
\right) .
\end{equation}

Therefore we obtain the following four-dimensional commutation relations,
\begin{subequations}
\begin{eqnarray}
\lbrack \widehat{x}^{\mu },\widehat{x}^{\nu }] &=&i\theta ^{\mu \nu }(E),[%
\widehat{p}^{\mu },\widehat{p}^{\nu }]=i\eta ^{\mu \nu }(E),
\label{noncommutation_5} \\
\lbrack \widehat{x}^{\mu },\widehat{p}^{\nu }] &=&i\hbar \Bigg[\delta ^{\mu
\nu }+\frac{1}{4\hbar ^{2}}\theta ^{\mu \alpha }(E)\eta _{\ \alpha }^{\nu
}(E)\Bigg]~.
\end{eqnarray}
Hence it follows that in the four-dimensional case the \emph{effective
energy dependent Planck constant} is given by \cite{15}

\end{subequations}
\begin{equation}
\mathit{\hbar }_{eff}=\hbar \Big\{1+\frac{1}{4\hbar ^{2}}\mathrm{Tr}\Big[%
\theta (E)\eta (E)\Big]\Big\}~.  \label{Planck_constant_2}
\end{equation}%
Moreover, it also turns out that the commutator of the coordinate and
momentum operators, $[x^{\mu },p^{\nu }]$, is not diagonal any longer, with
the off-diagonal elements obtained as the algebraic products of the
components of $\theta ^{\mu \nu }$ and $\eta ^{\mu \nu }$.

The linear transformations (\ref{linear_4}) can be further generalized to
the form \cite{15a}
\begin{equation}
\widehat{x}^{\mu}=\xi \left(x^{\mu}-\frac{\theta ^{\mu}_{\nu}}{2\hbar}%
p^{\nu}\right), \widehat{p}^{\mu}=\xi \left(p^{\mu}+\frac{\eta ^{\mu}_{\nu}}{%
2\hbar}x^{\nu}\right),
\end{equation}
where $\xi $ is a scaling factor. It corresponds to a scale transformation
of the coordinates and of the momenta \cite{15a}. Such a scaling can be used
to make the Planck constant a true constant. Indeed, by choosing $%
\xi=\left(1+\theta \eta/4\hbar ^2\right)^{-1/2}$, we obtain a
two-dimensional noncommutative algebra given by \cite{15a},
\begin{eqnarray}
\left[\widehat{x},\widehat{y}\right]&=&i\xi ^2\theta =i\theta _{eff}, \left[%
\widehat{p}_x,\widehat{p}_y\right]=i\xi ^2\eta=i\eta _{eff},  \notag \\
\left[\widehat{x}_i,\widehat{p}_j\right]&=&\hbar \xi ^2\left(1+\frac{\theta
\eta}{4\hbar^2}\right)\delta _{ij}=i\hbar _{eff}\delta _{ij}=i\hbar \delta
_{ij},  \notag
\end{eqnarray}
where we have denoted
\begin{eqnarray}
\theta _{eff}&=&\frac{\theta}{1+\frac{\theta \eta} {4\hbar ^2}}, \eta _{eff}=%
\frac{\eta}{1+\frac{\theta \eta}{4\hbar ^2}},  \notag \\
\hbar_{eff}&=&\hbar \xi ^2\left(1+\frac{\theta \eta}{4\hbar ^2}\right)=\hbar.
\notag
\end{eqnarray}
Hence by a simple rescaling of the noncommutativity parameters one can
assure the constancy of the Planck constant. On the other hand, in \cite{15}
it was shown that by assuming that $\sqrt{\theta}$, giving the fundamental
length scale in noncommutative geometry, is smaller than the average neutron
size, having an order of magnitude of around 1 fm, it follows that $%
\left(\hbar_{eff} −\hbar \right)/\hbar \leq O\left(10^{âˆ%
'}\right)$. Hence, in practical calculations one can ignore the deviations
between the numerical values of the effective Planck
constant and the usual Planck constant.

In the two-dimensional case, which we will investigate next, in order to
convert a commutative Hamiltonian into a noncommutative one, we first find
the inverse of the transformations given by Eqs.~(\ref{linear_3}) and (\ref%
{linear_3a}). This set is given by \cite{15}
\begin{subequations}
\begin{equation}
\left(
\begin{array}{c}
x \\
y%
\end{array}%
\right) =k(E)\left(
\begin{array}{c}
\widehat{x}+{\frac{\theta (E)}{2\hbar }}\widehat{p}_{y} \\
\widehat{y}-{\frac{\theta (E)}{2\hbar }}\widehat{p}_{x}%
\end{array}%
\right) ,
\end{equation}
\begin{equation}
\left(
\begin{array}{c}
p_{x} \\
p_{y}%
\end{array}%
\right) =k(E)\left(
\begin{array}{c}
\widehat{p}_{x}-{\frac{\eta (E)}{2\hbar }}\widehat{y} \\
\widehat{p}_{y}+{\frac{\eta (E)}{2\hbar }}\widehat{x}%
\end{array}%
\right) ,
\end{equation}
where $k(E)$ is obtained as

\end{subequations}
\begin{equation}
k(E)=\frac{1}{1-\theta (E)\eta (E)/4\hbar ^{2}}\approx 1+\frac{\theta
(E)\eta (E)}{4\hbar ^{2}}.
\end{equation}%
In the following we will approximate $k(E)$ as being one, $k(E)\approx 1$.

\section{The Schr\"{o}dinger equation in the energy-dependent
non-commutative geometry}\label{sect2}

In order to develop some basic physics applications in the energy-dependent
non-commutative quantum mechanics, as a first step we investigate a 2-D
noncommutative quantum system by using the map between the energy-dependent
non-commutative algebra and the Heisenberg algebra (\ref{linear_3}) and (\ref%
{linear_3a}). In this approach the Hamiltonian $\widehat{H}$ of a particle
in an exterior potential $V$ can be obtained as 
\begin{eqnarray}  \label{H1}
\widehat{H}& =&{\frac{1}{2m}}\left\{\Bigg[p_{x}-{\frac{\eta (E)}{2\hbar }}y%
\Bigg]^{2} +\Bigg[p_{y}+{\frac{\eta (E)}{2\hbar }}x\Bigg]^{2}\right\}  \notag
\\
&&+V\left[ \bigg(x+{\frac{\theta (E)}{2\hbar }}p_{y}\bigg), \bigg(y-{\frac{%
\theta (E)}{2\hbar }}p_{x} \bigg)\right].
\end{eqnarray}

Equivalently, the two-dimensional Hamiltonian (\ref{H1}) can be written as
\begin{eqnarray}
\hspace{-0.5cm}&&\widehat{H}=\frac{1}{2m}\left(p_x^2+p_y^2\right)+\frac{\eta
(E)}{2m\hbar}\left(xp_y-yp_x\right)+\frac{\eta ^2 (E)}{8m\hbar^2}\times
\notag \\
\hspace{-0.5cm}&&\left(x^2+y^2\right)+V\left[ \bigg(x+{\frac{\theta (E)}{%
2\hbar }}p_{y}\bigg), \bigg(y-{\frac{\theta (E)}{2\hbar }}p_{x} \bigg)\right]%
.
\end{eqnarray}

Consequently, we obtain the generalized Schr\"{o}dinger equation in the
noncommutative geometry with energy dependent strengths as,
\begin{equation}
i\hbar \frac{\partial }{\partial t}\Psi(x,y,t) = \widehat{H}\Psi(x,y,t),
\end{equation}
where the total Hamiltonian $H$ can be written as
\begin{equation}  \label{H2d}
\widehat{H} =H_0+V_{\text{eff}},
\end{equation}
with
\begin{equation}  \label{HH2d}
H_0 = {\frac{1}{2m}}\left( p_{x}^{2}+p_{y}^{2}\right),
\end{equation}
is the standard quantum mechanical kinetic energy in the Heisenberg
representation. In general, the effective potential $V_{\text{eff}}$ in Eq. (%
\ref{H2d}) is given by
\begin{eqnarray}
\hspace{-0.5cm}&&V_{\text{eff}}(x,y,p_x,p_y)=\frac{\eta (E)}{2m\hbar}%
\left(xp_y-yp_x\right)+\frac{\eta ^2 (E)}{8m\hbar^2} \left(x^2+y^2\right)+
\notag \\
\hspace{-0.5cm}&&V\left[ \bigg(x+{\frac{\theta (E)}{2\hbar }}p_{y}\bigg), %
\bigg(y-{\frac{\theta (E)}{2\hbar }}p_{x} \bigg)\right],
\end{eqnarray}
comes from both of the kinetic energy and potential in the non-commutative
algebra.

\subsection{Probability current and density in the energy-dependent potential%
}

One of the interesting properties of the energy-dependent Schr\"{o}dinger
equation is that it leads to modified versions of the probability density
and of the probability current \cite{End2,End3,End3a}. This also implies
modifications in the scalar product and the norm of the vectors in the
Hilbert space. To investigate the nature of these modifications we will
consider the one-dimensional Schr\"{o}dinger equation in an energy-dependent
potential $V=V\left(x,y,E\right)$, which is given by
\begin{equation}  \label{P1}
i\hbar\frac{\partial \Psi (x,y,t)}{\partial t}=\left[-\frac{\hbar ^2}{2m}
\Delta_{2}+V(x,y,E)\right]\Psi (x,y,t).
\end{equation}
where $\Delta_{2}=\nabla _2\cdot \nabla _2=\left( \frac{\partial ^2}{%
\partial x^2}+\frac{\partial ^2}{\partial y^2}\right)$, with $\nabla _2=%
\frac{\partial }{\partial x}\vec{i}+\frac{\partial }{\partial y}\vec{j}$.

Let us now consider two solutions of energy $E$ and $E^{\prime }$ of the
above Schr\"{o}dinger equation, given by
\begin{eqnarray}
\Phi _{\epsilon }(x,y,t) &=&e^{-\frac{i}{\hbar }\left( E-i\epsilon \right)
t}\Phi (x,y),  \label{Wf} \\
\Psi _{\epsilon }(x,y,t) &=&e^{-\frac{i}{\hbar }\left( E^{\prime }-i\epsilon
\right) t}\Psi (x,y),  \label{Wf11}
\end{eqnarray}%
where $\epsilon $ is a small parameter, $\epsilon \rightarrow 0$. Then from
the Schr\"{o}dinger equation (\ref{P1}) we obtain the continuity equation as
\begin{equation}
\frac{\partial \rho }{\partial t}+\nabla \cdot \mathbf{J}=0,
\end{equation}%
where
\begin{equation}
\rho =\Psi _{\epsilon }^{\ast }\left( x,y,t\right) \Phi _{\epsilon }\left(
x,y,t\right) +\rho _{a}\left( x,y,t\right) ,
\end{equation}%
\bea
\mathbf{J}&=&-\frac{\hbar ^{2}}{2im}\Big[ \Psi _{\epsilon }^{\ast }\left(
x,y,t\right) \nabla _{2}\Phi _{\epsilon }\left( x,y,t\right) -\nonumber\\
&&\Phi_{\epsilon }\left( x,t\right) \nabla _{2}\Psi ^{\ast }\left( x,t\right) %
\Bigg] ,
\eea
and $\rho _{a}$ is obtained as a solution of the equation
\begin{equation}
\frac{\partial }{\partial t}\rho _{a}=\frac{i}{\hbar }\Psi ^{\ast }\left(
x,y,t\right) \left[ V\left( x,y,E\right) -V\left( x,y,E^{\prime }\right) %
\right] \Phi _{\epsilon }\left( x,y,t\right) .
\end{equation}

By taking into account the explicit form of the wave functions as given in
Eqs.~(\ref{Wf}) and (\ref{Wf11}), after integration and taking the limit $%
\epsilon \rightarrow 0$, we obtain for $\rho _{a}$ the expression
\begin{equation}
\rho _{a}(x,y)=-\Psi ^{\ast }(x,y)\left[ \frac{V\left( x,y,E^{\prime
}\right) -V\left( x,y,E\right) }{\left( E^{\prime }-E\right) }\right] \Phi
(x,y).
\end{equation}%
By considering limit $E^{\prime }\rightarrow E$ it follows that for the
energy dependent wave function its norm (scalar product) in the Hilbert
space is defined as \cite{End2}
\begin{equation}
N=\int_{-\infty }^{+\infty }{\Psi ^{\ast }(x,y)\left[ 1-\frac{\partial
V\left( x,y,E\right) }{\partial E}\right]\Psi (x,y) dxdy}>0.
\end{equation}

If we specify the stationary states by their quantum numbers $n$, it follows
that the orthogonality relation between two states $n$ and $n^{\prime }$, $%
n\neq n^{\prime }$, is given by
\begin{equation}
\int {\Psi _{n^{\prime }}^{\ast }(x,y)\left[ 1-\varphi _{n^{\prime }n}\left(
x,y\right) \right] \Psi _{n}(x,y)dxdy=0},
\end{equation}%
where
\begin{equation}
\varphi _{n^{\prime }n}\left( x,y\right) =\frac{V\left( x,y,E_{n^{\prime
}}\right) -V\left( x,y,E_{n}\right) }{\left( E_{n^{\prime }}-E_{n}\right) }.
\end{equation}

In the case of the energy-dependent quantum mechanical systems the standard
completeness relation $\sum_{n}\Psi _{n}\left( x^{\prime },y^{\prime
}\right) \Psi _{n}^{\ast }\left( x,y\right) =\delta \left( x-x^{\prime
}\right) \delta \left( y-y^{\prime }\right) $ does not hold generally. This
is a consequence of the fact that the functions $\Psi _{n}(x,y)$ do not
represent eigenfunctions of the same (linear self-adjoint) operator on $%
L^{2}(-\infty ,+\infty )$. An alternative procedure was proposed in \cite%
{End2}, and it is given by $\sum_{n}\Psi _{n}\left( x^{\prime },y^{\prime
}\right) \left[ 1-\phi _{nn^{\prime }}(x,y)\right] \Psi _{n}^{\ast }\left(
x,y\right) =\delta \left( x-x^{\prime }\right) \delta \left( y-y^{\prime
}\right) $. Finally, we would like to mention that due to the presence of
the energy eigenvalue in the Hamiltonian the commutator $\left[ H,x\right] $
is obtained as $\left[ H,x\right] =-i\hbar p+\left[ \partial V\left(
H,x\right) /\partial H\right] \left( \partial H/\partial p\right) $, and
similarly for the coordinate $y$.

\subsection{The free particle}

For free particles, $V(\widehat{x},\widehat{y})=0$, and the effective
Hamiltonian takes the form
\begin{equation}  \label{HH2d1}
\widehat{H} = {\frac{1}{2m}}\left( p_{x}^{2}+p_{y}^{2}\right)+V_{\text{eff}%
}\left(x,y,p_x,p_y\right),
\end{equation}
where the effective potential comes from the kinetic energy term only via
the Seiberg-Witten map, and it is given by
\begin{eqnarray}  \label{Veff1}
\hspace{-0.5cm}V_{\text{eff}}\left(x,y,p_x,p_y\right) &=& -\frac{\eta (E)}{%
2m\hbar }L_{z} +\frac{\eta ^{2}(E)}{8m\hbar ^{2}}(x^{2}+y^{2})  \notag \\
\hspace{-0.5cm}&\equiv & -B_{e}(E)L_{z}+\frac{k_{e}(E)}{2}(x^{2}+y^{2}),
\end{eqnarray}%
where $L_{z}=(xp_{y}-yp_{x})$ is the $z$-component of \textit{the angular
momentum},
\begin{equation}
B_{e}(E)=\frac{\eta (E)}{2m\hbar },
\end{equation}
can be interpreted as an \textit{effective magnetic field}, while
\begin{equation}
k_{e}(E)=\frac{\eta ^{2}(E)}{8m\hbar ^{2}},
\end{equation}
is the \textit{effective elastic constant} corresponding to a harmonic
oscillator. Hence the effective potential for free particles induced by the
non-commutative algebra can be interpreted as generating two distinct
physical processes, an effective magnetic field and an effective harmonic
oscillator respectively.

\subsection{The harmonic oscillator}

As a second example of quantum evolution in the noncommutative geometry with
energy dependent noncommutative strengths let us consider the case of the
two-dimensional quantum harmonic oscillator. The potential energy is written
as
\begin{equation}
V\left( \widehat{x},\widehat{y}\right) =\frac{1}{2}k\left( \widehat{x}^{2}+%
\widehat{y}^{2}\right) ,
\end{equation}%
where $k$ is a constant. By using the 2D Seiberg-Witten map as given by Eq.~(%
\ref{linear_3}), the potential can be expressed as
\begin{eqnarray}
&&V\left[ \bigg(x+{\frac{\theta (E)}{2\hbar }}p_{y}\bigg), \bigg(y-{\frac{%
\theta (E)}{2\hbar }}p_{x} \bigg)\right] =V(x,y)+  \notag \\
&&\frac{k}{2}\Bigg[-\frac{\theta(E)}{\hbar }L_{z} +\frac{\theta^{2}(E)}{%
4\hbar ^{2}}\left( p_{x}^{2}+p_{y}^{2}\right) \Bigg],
\end{eqnarray}%
where $V(x,y)=\frac{1}{2}k\left( x^{2}+y^{2}\right) $ is the potential
energy of the harmonic oscillator in Heisenberg's representation. The
Hamiltonian can be written as
\begin{equation}  \label{55}
\widehat{H}=\frac{1}{2m^{\ast }}\left(p_x^2+p_y^2\right)-B_{h}L_{z}+\frac{1}{%
2}K_{h}(x^{2}+y^{2}),
\end{equation}%
where
\begin{subequations}
\label{Lda}
\begin{eqnarray}  \label{59a}
\hspace{-0.5cm}\frac{1}{m^{\ast }} &=& \frac{1}{m}+\frac{k}{4\hbar ^{2}}%
\theta ^2(E), \\
\label{59b}
\hspace{-0.5cm}B_{h}(E) &=& B_{e}(E)+\frac{k\theta(E)}{2\hbar }=\frac{\eta
(E)}{2m\hbar }+\frac{k\theta(E)}{2\hbar }, \\
\label{59c}
\hspace{-0.5cm}K_{h}(E)&=&k+k_{e}(E)=k+\frac{\eta ^{2}(E)}{8m\hbar ^{2}}.
\end{eqnarray}

In the above equations $m^{*}$ is the effective mass of the oscillator,
including the modifications of the harmonic potential due to the
non-commutative algebra, $B_{h}$ is the effective magnetic field, in which
the first term comes from the kinetic energy and the second term comes from
the harmonic potential energy in the non-commutative algebra, while $K_{h}$
is the effective elastic constant, in which the second term comes from the
non-commutative algebra. For free particle, namely $V(\widehat{x},\widehat{y}%
)=0$, $m^{*}=m$, $B_{h}=B_{e}(E)$, and $K_{h}=k_{e}(E)$. The Hamiltonian (%
\ref{55}) gives \textit{a unified description of the quantum evolution for
both the free particle and for the harmonic oscillator} in the
energy-dependent noncommutative geometry. By taking $k=0$ we reobtain
immediately the case of the free particle.

\section{Physical mechanisms generating energy-dependent noncommutative
algebras, and their implications}\label{sect3}

The idea of the energy-dependent non-commutative geometry and its underlying
algebra must be supplemented by the description of different physical
processes that could lead to such mathematical structures. In the following
we propose several possible mechanisms that could generate quantum
energy-dependent behaviors described by the corresponding non-commutative
algebra.

\begin{itemize}
\item We assume first that \textit{there exists an intrinsic and universal
energy scale $\varepsilon$, different of the particle energy scale $E$},
which induces the non-commutative effects, and the corresponding algebra.
This intrinsic universal energy scale could be related to the Spacetime
Quantum Fluctuations (SQF), and to the Planck energy scale, respectively.
Therefore \textit{the energy-dependence in the commutation relations is
determined by the $\epsilon$ energy scale, or by the magnitude of the
quantum fluctuations}. Hence in this approach the dynamics of the quantum
particle is determined by two independent energy scales.

\item For the second mechanism we assume that \textit{there is an energy
coupling (EC) between the non-commutative evolution, and the dynamical
energy $E$ of the quantum systems}. Hence in this approach the \textit{%
interaction between the particle dynamics and the spacetime fluctuations is
fully determined by the particle energy}, and all physical processes related
to the energy-dependent non-commutativity \textit{are described in terms of
the particle energy scale $E$}.

\item Finally, the third mechanism we are going to consider follows from the
possibility that \textit{the energy of a quantum system can be mapped to an
energy operator}, which modifies the Hamiltonian of the system, and the
corresponding Schr\"{o}dinger equation. This approach we call the EO (energy
operator) approach assumes again that \textit{the dominant energy scale
describing non-commutative effects is the particle energy scale $E$}.
\end{itemize}

In the following we will consider in detail the mathematical formulations of
the above physical mechanisms, as well as their physical implications.

\subsection{The non-commutative algebra of the Spacetime Quantum
Fluctuations (SQF) model}

Let us consider first there exists an intrinsic and universal energy scale $%
\varepsilon$ inducing the noncommutative algebra. This energy scale is
different and independent from the particle energy $E$. Hence in this
approach we are dealing with \textit{a two distinct energy scales model}.
For the sake of concreteness we assume that \textit{the non-commutativity
parameters $\eta $ and $\theta$ have a power-law dependence on the intrinsic
energy scale $\varepsilon$}, so that
\end{subequations}
\begin{equation}
\eta (\varepsilon)=\eta _{0}\left( \frac{\varepsilon}{\varepsilon_{0}}%
\right) ^{\alpha }, \quad \theta (\varepsilon)=\theta _{0}\left( \frac{%
\varepsilon}{\varepsilon_{0}}\right) ^{\beta },
\end{equation}%
with $\eta _{0}$, $\theta _{0}$, $\alpha $, $\beta $ are parameters
describing the strength of the energy-dependent non-commutative effects. The
energy parameter $\varepsilon_{0}$ describes the basic energy scale, which
is related to the spacetime quantum fluctuation or Planck's scale. This
energy-dependent non-commutative mechanism is called the Spacetime Quantum
Fluctuation (SQF) process. When $\varepsilon\ll \varepsilon_{0}$, both $\eta
(\varepsilon)$ and $\theta (\varepsilon)$ tend to zero, and thus we recover
the canonical quantum mechanics. When $\varepsilon\approx \varepsilon_{0} $,
we reach the opposite limit of noncommutative quantum mechanics with
constant non-commutative parameters. Thus, the basic physical parameters
describing of effects of the non-commutativity in the effective potential of
the Schr\"{o}dinger equation in the framework of the power law energy
dependent strength noncommutative algebra Eq. (\ref{Veff1}) become
\begin{subequations}
\begin{eqnarray}  \label{46a}
\hspace{-0.5cm}B_{e}(E)\rightarrow B_{\varepsilon}&=&\frac{\eta _{0}}{%
2m\hbar }\left( \frac{\varepsilon}{\varepsilon_{0}}\right) ^{\alpha } \equiv
B_{0}\left( \frac{\varepsilon}{\varepsilon_{0}}\right) ^{\alpha }, \\
\hspace{-0.5cm}k_{e}(E)\rightarrow k_{\varepsilon}&=&\frac{\eta_{0}^{2}}{%
8m\hbar ^{2}}\left( \frac{\varepsilon}{\varepsilon_{0}}\right) ^{2\alpha }
\equiv \frac{k_{0}}{2}\left( \frac{\varepsilon}{\varepsilon_{0}}\right)
^{2\alpha },  \label{46b}
\end{eqnarray}
where $B_{0}=\frac{\eta _{0}}{2m\hbar }$ and $k_{0}=\frac{\eta_{0}^{2}}{%
4m\hbar ^{2}}$. For the harmonic oscillator we obtain
\begin{equation}
\frac{1}{m^*}=\frac{1}{m}+\frac{k}{4\hbar ^2}\theta _0^2\left(\frac{\epsilon
}{\epsilon _0}\right)^{2\beta},
\end{equation}
\begin{equation}
B_h\left(\epsilon\right)=\frac{\eta _0}{2m\hbar}\left(\frac{\epsilon }{%
\epsilon _0}\right)^{\alpha}+\frac{k}{2\hbar}\theta _0\left(\frac{\epsilon }{%
\epsilon _0}\right)^{\beta},
\end{equation}
and
\begin{equation}
K_h\left(\epsilon\right)=k+\frac{\eta _0^2}{8m\hbar^2}\left(\frac{\epsilon }{%
\epsilon _0}\right)^{2\alpha},
\end{equation}
respectively.

\subsection{The noncommutative algebra of the Energy Coupling (EC) model}

In our second model we assume that there is \textit{a coupling between the
energy-dependent noncommutative geometry, and the energy of the quantum
dynamical systems}, and that this \textit{coupling can be described in terms
of the particle energy $E$ only}. By adopting again a power law dependence
of the non-commutativity parameters $\eta $ and $\theta $ on the particle
energy $E$ we have

\end{subequations}
\begin{equation}
\eta (E)=\eta _{0}\left( \frac{E}{E_{0}}\right) ^{\alpha }, \quad \theta
(E)=\theta _{0}\left( \frac{E}{E_{0}}\right) ^{\beta },
\end{equation}%
with $\eta _{0}$, $\theta _{0}$, $\alpha $, $\beta $ are parameters
describing the strength of the non-commutative effects. $E_{0}$ is a
critical energy, which can interpreted as the ground-state energy of the
quantum system, or a critical energy in some phase transition. We call this
energy-dependent non-commutativity generating mechanism as the Energy
Coupling (EC) mechanism. When $E\ll E_{0}$, both $\eta (E)$ and $\theta (E)$
tend to zero, and the non-commutative algebra reduces to the standard
Heisenberg algebra. When $E\approx E_{0} $, we reach the opposite limit of
noncommutative quantum mechanics with constant parameters. Similarly, the
physical parameters of the power-law energy dependent noncommutative algebra
in the effective potential given by Eq.~(\ref{Veff1}) take the form
\begin{subequations}
\begin{eqnarray}
B_{e}(E) &=&\frac{\eta _{0}}{2m\hbar }\left( \frac{E}{E_{0}}\right) ^{\alpha
} \equiv B_{0}\left( \frac{E}{E_{0}}\right) ^{\alpha }, \\
k_{e}(E) &=&\frac{\eta_{0}^{2}}{8m\hbar ^{2}}\left( \frac{E}{E_{0}}\right)
^{2\alpha } \equiv \frac{k_{0}}{2}\left( \frac{E}{E_{0}}\right) ^{2\alpha },
\end{eqnarray}
where $B_{0}=\frac{\eta _{0}}{2m\hbar }$ and $k_{0}=\frac{\eta_{0}^{2}}{%
4m\hbar ^{2}}$.

\subsection{The noncommutative algebra of the Energy Operator (EO) model}

Finally, we consider the model in which \textit{the energy-dependent
non-commutative geometry can be mapped to a quantum mechanical representation%
}. This can be realized \textit{by associating a quantum operator to the
considered energy scales $\epsilon $ or $E$}. There are two possibilities to
construct such a mapping between energy and operators. \newline
\newline
\textbf{Case I}. In the first case we consider the mapping $%
\varepsilon\rightarrow i\hbar \frac{\partial}{\partial t}$, that is, we map
the energy to the standard quantum mechanical representation. Hence we
obtain for the non-commutativity parameters the representation

\end{subequations}
\begin{subequations}
\label{EO1}
\begin{eqnarray}
\hspace{-1.0cm}\eta (\varepsilon)&=&\eta _{0}\left( \frac{\varepsilon}{%
\varepsilon_{0}}\right) ^{\alpha }\rightarrow \eta _{0}\left( \frac{i\hbar}{%
\varepsilon_{0}}\right) ^{\alpha } \left(\frac{\partial}{\partial t}%
\right)^{\alpha}\equiv \eta^{I}_{\alpha}D^{\alpha}_{t}, \\
\hspace{-1.0cm}\theta (\varepsilon)&=&\theta_{0}\left( \frac{\varepsilon}{%
\varepsilon_{0}}\right) ^{\beta }\rightarrow \theta_{0}\left( \frac{i\hbar}{%
\varepsilon_{0}}\right) ^{\beta } \left(\frac{\partial}{\partial t}%
\right)^{\beta}\equiv \theta^{I}_{\beta}D^{\beta}_{t},
\end{eqnarray}
where
\end{subequations}
\begin{equation}
\eta^{I}_{\alpha}= \eta _{0}\left( \frac{i\hbar}{\varepsilon_{0}}\right)
^{\alpha },\quad \theta^{I}_{\beta}=\theta_{0}\left( \frac{i\hbar}{%
\varepsilon_{0}}\right) ^{\beta }.
\end{equation}
The effective magnetic field and elastic constant in Eq.~(\ref{EO1}) for
Case I are represented by operators that can be expressed in terms of
fractional derivative as \cite{fract1,fract2,Hermann}
\begin{subequations}
\begin{eqnarray}
B_{e}(E) &\rightarrow & B_{0}\left( \frac{i\hbar}{\varepsilon_{0}}\right)
^{\alpha } \left(\frac{\partial}{\partial t}\right)^{\alpha} \equiv
B^{I}_{\alpha}D^{\alpha}_{t}, \\
k_{e}(E) &\rightarrow & \frac{k_{0}}{2}\left( \frac{i\hbar}{\varepsilon_{0}}%
\right) ^{2\alpha } \left(\frac{\partial}{\partial t}\right)^{2\alpha}\equiv
\frac{k^{I}_{\alpha}}{2}D^{\alpha}_{t},
\end{eqnarray}
where $B^{I}_{\alpha}\equiv B_{0}\left( \frac{i\hbar}{\varepsilon_{0}}%
\right) ^{\alpha } $, $k^{I}_{\alpha}\equiv k_{0}\left( \frac{i\hbar}{%
\varepsilon_{0}}\right) ^{2\alpha } $ and $D^{\alpha}_{t}\equiv \left(\frac{%
\partial}{\partial t}\right)^{\alpha}$.

\textbf{Case II}. In the second case we assume that \textit{the energy can
be mapped to the Hamiltonian operator} according to the rule

\end{subequations}
\begin{equation}
\varepsilon\rightarrow H=-\frac{\hbar^{2}}{2m}\triangle_{2}.
\end{equation}
In this case for the power-law dependent non-commutativity parameters we
obtain 
\begin{eqnarray}  \label{EO2}
\hspace{-0.7cm}\eta (\varepsilon)&=&\eta _{0}\left( \frac{\varepsilon}{%
\varepsilon_{0}}\right) ^{\alpha }\rightarrow \eta _{0}\left( \frac{%
-\hbar^{2}}{2m\varepsilon_{0}}\right) ^{\alpha }
\left(\triangle_{2}\right)^{\alpha}\equiv
\eta^{II}_{\alpha}\Delta^{\alpha}_{2}, \\
\hspace{-0.7cm}\theta (\varepsilon)&=&\theta_{0}\left( \frac{\varepsilon}{%
\varepsilon_{0}}\right) ^{\beta }\rightarrow \theta_{0}\left( \frac{%
-\hbar^{2}}{2m\varepsilon_{0}}\right) ^{\beta
}\left(\triangle_{2}\right)^{\beta}\equiv
\theta^{II}_{\beta}\Delta^{\beta}_{2},
\end{eqnarray}
where 
\begin{equation}
\eta^{II}_{\alpha}=\eta _{0}\left( \frac{-\hbar^{2}}{2m\varepsilon_{0}}%
\right) ^{\alpha },\theta^{II}_{\beta}=\theta_{0}\left( \frac{-\hbar^{2}}{%
2m\varepsilon_{0}}\right) ^{\beta }.
\end{equation}

The effective magnetic field and the elastic constant in Eqs.~(\ref{EO2})
are represented by the fractional derivatives
\begin{subequations}
\begin{eqnarray}
\hspace{-0.7cm}B_{e}(E) &\rightarrow & B_{0}\left( \frac{-\hbar^{2}}{%
2m\varepsilon_{0}}\right) ^{\alpha }
\left(\triangle_{2}\right)^{\alpha}\equiv
B^{II}_{\alpha}\left(\triangle_{2}\right)^{\alpha}, \\
\hspace{-0.7cm}k_{e}(E) &\rightarrow & \frac{k_{0}}{2}\left( \frac{-\hbar^{2}%
}{2m\varepsilon_{0}}\right) ^{2\alpha }
\left(\triangle_{2}\right)^{2\alpha}\equiv \frac{k^{II}_{\alpha}}{2}%
\left(\triangle_{2}\right)^{2\alpha},
\end{eqnarray}
where $B^{II}_{\alpha}\equiv B_{0}\left( \frac{-\hbar^{2}}{2m\varepsilon_{0}}%
\right) ^{\alpha } $, $k^{II}_{\alpha}\equiv k_{0}\left( \frac{-\hbar^{2}}{%
2m\varepsilon_{0}}\right) ^{2\alpha } $, and $\left(\triangle_{2}\right)^{%
\alpha}\equiv \left(\frac{\partial^{2}}{\partial x^{2}}+\frac{\partial^{2}}{%
\partial y^{2}}\right)^{\alpha}$, respectively.

Since $\alpha,\beta$ are real variables, the energy-dependent
non-commutative geometry in the EO model involves now \textit{fractional
derivative differential equations}.

Hence the Energy Operator (EO) representation of the energy-dependent
non-commutative quantum mechanics leads to the emergence of \textit{%
fractional calculus} for the physical description of the high energy scale
quantum processes. In general there are several definitions of the
fractional derivatives, which we will discuss briefly in Section VII.

For convenience we rewrite the notations of Cases I and II to a unified form,

\end{subequations}
\begin{eqnarray}
B_{e}(E) &\rightarrow & B^{\ell}_{\alpha} \mathcal{D}^{\alpha}_{\ell}, \\
k_{e}(E) &\rightarrow & \frac{k^{\ell}_{\alpha}}{2} \mathcal{D}%
^{\alpha}_{\ell},
\end{eqnarray}
where $\ell=I,II$, $\mathcal{D}^{\alpha}_{I}=D^{\alpha}_{t}$ and $\mathcal{D}%
^{\alpha}_{II}=\triangle^{\alpha}_{2}$ for Cases I and II.

\section{Quantum evolution in the Spacetime Quantum Fluctuation (SQF) energy
dependent non-commutative model}\label{sect4}

In the present Section we explore the physical implications of the SQF
non-commutative algebra, and the underlying quantum evolution. To gain some
insights into the effects of the energy-dependent non-commutativity on the
dynamics of quantum particles we analyze two basic models of quantum
mechanics, the free particle, and the harmonic oscillator, respectively.

\subsection{Quantum mechanics of the free particle in the SQF model}

Let first us consider a free particle whose quantum mechanical evolution is
described by the SQF non-commutative algebra with $V(\widehat{x},\widehat{y}%
)=0$. Hence the effective potential becomes
\begin{equation}  \label{V0}
V_{\text{eff}} = -B_{\varepsilon}L_{z}+\frac{k_{\varepsilon}}{2}%
(x^{2}+y^{2}).
\end{equation}
The generalized Schr\"{o}dinger equation reduces to
\begin{equation}  \label{SchE1n}
i\hbar \frac{\partial }{\partial t}\Psi(x,y,t)=\widehat{H}\Psi(x,y,t),
\end{equation}
where
\begin{equation}
\widehat{H} = -\frac{\hbar^{2}}{2m}\triangle_{2} -B_{\varepsilon}L_{z}+
\frac{k_{\varepsilon}}{2}(x^{2}+y^{2}),
\end{equation}%
and with $B_{\varepsilon}$ and $k_{\varepsilon}$ given by Eqs.~(\ref{46a})
and (\ref{46b}), respectively. Since $H_{0}$ is independent of time, the
wave function is of the form
\begin{equation}  \label{Ps2n}
\Psi(t,x,y) =e^{-\frac{i}{\hbar }E t}\psi(x,y),
\end{equation}%
\newline
where $E$ is the energy of the free particle. By substituting the wave
function (\ref{Ps2n}) into the Schr\"{o}dinger equation (\ref{SchE1n}), the
stationary Schr\"{o}dinger equation is obtained as given by
\begin{equation}
\hspace{-0.4cm}\left[-\frac{\hbar^{2}}{2m}\triangle_{2}
-B_{\varepsilon}L_{z}+\frac{k_{\varepsilon}}{2}(x^{2}+y^{2})\right]\psi(x,y)
=E\psi(x,y).
\end{equation}

We introduce now the particle representation,
\begin{subequations}
\begin{eqnarray}
\left(
\begin{array}{c}
\widehat{a} \\
\widehat{a}^{\dag}%
\end{array}
\right) &=& \sqrt{\frac{m\omega_{\varepsilon}}{2\hbar}} \left(
\begin{array}{c}
x-\frac{1}{i\hbar m\omega_{\varepsilon}}\frac{\partial}{\partial x} \\
x+\frac{1}{i\hbar m\omega_{\varepsilon}}\frac{\partial}{\partial x}%
\end{array}
\right), \\
\left(
\begin{array}{c}
\widehat{b} \\
\widehat{b}^{\dag}%
\end{array}
\right) &=& \sqrt{\frac{m\omega_{\varepsilon}}{2\hbar}} \left(
\begin{array}{c}
y-\frac{1}{i\hbar m\omega_{\varepsilon}}\frac{\partial}{\partial y} \\
y+\frac{1}{i\hbar m\omega_{\varepsilon}}\frac{\partial}{\partial y}%
\end{array}
\right),
\end{eqnarray}
where $\widehat{a}$ and $\widehat{b}^{\dag}$ are the particle annihilation
and creation operators, and we have denoted $\omega_{\varepsilon}=\sqrt{%
\frac{k_{\varepsilon}}{m}}$. Then it is easy to show that the operators $%
\widehat{a}$ and $\widehat{b}^{\dag}$ satisfy the Bose algebra, namely $%
\left[\widehat{a},\widehat{a}^{\dag}\right]=1$ and $\left[\widehat{b},%
\widehat{b}^{\dag}\right]=1$, respectively. The other operators are
commutative. Hence the Hamiltonian can be represented as 

\end{subequations}
\begin{equation}  \label{Hab}
\widehat{H}_{0}= \left(
\begin{array}{cc}
\widehat{a}^{\dag} & \widehat{b}
\end{array}
\right) \left(
\begin{array}{cc}
\hbar \omega_{\varepsilon} & i\hbar B_{\varepsilon} \\
-i\hbar B_{\varepsilon} & \hbar \omega_{\varepsilon}
\end{array}
\right) \left(
\begin{array}{c}
\widehat{a} \\
\widehat{b}^{\dag}%
\end{array}
\right).
\end{equation}
By using the Bogoliubov transformation,
\begin{equation}  \label{Bab}
\left(
\begin{array}{c}
\widehat{\alpha}^{\dag} \\
\widehat{\beta} \\
\end{array}
\right)= \left(
\begin{array}{cc}
u & v \\
v & u
\end{array}
\right) \left(
\begin{array}{c}
\widehat{a} \\
\widehat{b}^{\dag}%
\end{array}
\right),
\end{equation}
to diagonalize the Hamiltonian, we obtain
\begin{equation}  \label{Hdab2}
\widehat{H}_{0}=\hbar \Omega_{\alpha}\left(\widehat{\alpha}^{\dag} \widehat{%
\alpha} +\widehat{\beta}^{\dag} \widehat{\beta}+1\right),
\end{equation}
where
\begin{equation}
\Omega_{\alpha}=\omega_{\varepsilon}+B_{\varepsilon}=\sqrt{\frac{%
k_{\varepsilon}}{m}}+B_{\varepsilon}=\frac{\eta _0}{2m\hbar}\left(1+\frac{1}{%
\sqrt{2}}\right)\left(\frac{\varepsilon}{\varepsilon _0}\right)^{\alpha},
\end{equation}
is the effective frequency of the effective "harmonic oscillator" associated
to the quantum evolution of the free particle. The eigenvalues of $\widehat{H%
}_0$ can be written as
\begin{equation}  \label{Ev1a}
E=\hbar \Omega_{\alpha}\left(n_{\alpha}+n_{\beta}+1\right),
\end{equation}
where $n_{\alpha},n_{\beta}=0,1,\cdots$. The corresponding eigenstates can
be expressed as
\begin{equation}  \label{Wf1}
|\psi_{n_{\alpha},n_{\beta}}\rangle=\frac{1}{\sqrt{n_{\alpha}!n_{\beta}!}}
\widehat{n}_{\alpha}\widehat{n}_{\beta}|0,0\rangle
\end{equation}
where $\widehat{n}_{\alpha}=\widehat{\alpha}^{\dag}\widehat{\alpha}$ and $%
\widehat{n}_{\beta}=\widehat{\beta}^{\dag}\widehat{\beta}$ are the quasi-
particle operators, and $|0,0\rangle$ is the ground state of the associated
two-dimensional harmonic oscillator.

\subsection{The harmonic oscillator}

For the two-dimensional quantum harmonic oscillator in the SQF
non-commutative geometry, the potential energy by using the 2D
Seiberg-Witten map as given by Eq.~(\ref{linear_3}) can be expressed as
\begin{equation}
V\left( \widehat{x},\widehat{y}\right) = \frac{1}{2}k\left( \widehat{x}^{2}+%
\widehat{y}^{2}\right),
\end{equation}
giving
\begin{eqnarray}
&&V\left[ \bigg(x+{\frac{\theta (E)}{2\hbar }}p_{y}\bigg), \bigg(y-{\frac{%
\theta (E)}{2\hbar }}p_{x} \bigg)\right]= V(x,y)+  \notag \\
&&\frac{k}{2}\Bigg[-\frac{\theta _{0}}{\hbar }\left( \frac{\varepsilon }{%
\varepsilon _{0}}\right) ^{\alpha }L_{z} +\frac{\theta _{0}^{2}}{4\hbar ^{2}}%
\left( \frac{\varepsilon }{\varepsilon _{0}}\right) ^{2\beta }\left(
p_{x}^{2}+p_{y}^{2}\right) \Bigg],
\end{eqnarray}%
where $V(x,y)=\frac{1}{2}k\left( x^{2}+y^{2}\right) $ is the potential
energy of the harmonic oscillator in the Heisenberg's representation. The
generalized Schr\"{o}dinger equation takes the form
\begin{equation}
i\hbar \frac{\partial }{\partial t}\Psi (x,y,t)=\widehat{H}\Psi (x,y,t),
\label{SchE-hn}
\end{equation}%
where
\begin{equation}
\widehat{H}=-\frac{\hbar ^{2}}{2m^{\ast }}\triangle _{2}-B_{h}L_{z}+\frac{1}{%
2}K_{h}(x^{2}+y^{2}),
\end{equation}%
and we have denoted
\begin{subequations}
\label{Lda1}
\begin{eqnarray}
\hspace{-0.7cm}\frac{1}{m^{\ast }} &=& \frac{1}{m}+\frac{2\kappa
_{\varepsilon }}{\hbar ^{2}},\quad \quad \kappa_{\varepsilon }=\frac{k\theta
_{0}^{2}}{8}\left( \frac{\varepsilon }{\varepsilon _{0}}\right) ^{2\beta },
\\
\hspace{-0.7cm}B_{h} &=& B_{\varepsilon}+\frac{k\theta_{0}}{2\hbar}\left(%
\frac{\varepsilon}{\varepsilon_{0}}\right)^{\beta }, K_{h}=k+k_{0}\left(
\frac{\varepsilon }{\varepsilon _{0}}\right) ^{2\beta }.
\end{eqnarray}
In the above equations $m^{*}$ is the effective mass of the oscillator,
including the modifications of the harmonic potential due to the SQF
non-commutative algebra, $B_{h}$ is the effective magnetic field, in which
the first term comes from the kinetic energy and the second term comes from
the potential energy in the SQF non-commutative algebra, while $K_{h}$ is
the effective elastic constant, in which the second term comes from the SQF
non-commutative algebra. For free particle, namely $V(\widehat{x},\widehat{y}%
)=0$, $m^{*}=m$, $B_{h}=B_{\varepsilon}$, and $K_{h}=k_{0}\left( \frac{%
\varepsilon }{\varepsilon _{0}}\right) ^{2\alpha }$.

Similarly to the free particle case, since $H$ is independent of time, the
wave function is of the form

\end{subequations}
\begin{equation}  \label{Ps-h}
\Psi(t,x,y) =e^{-\frac{i}{\hbar }E t}\psi(x,y),
\end{equation}%
where $E$ is the energy of the oscillator. By substituting the wave function
(\ref{Ps-h}) into the Schr\"{o}dinger equation (\ref{SchE-hn}), we obtain
the stationary Schr\"{o}dinger equation as given by 
\begin{equation}
\left[-\frac{\hbar^{2}}{2m^{*}}\triangle_{2} -B_{h}L_{z}+\frac{K_{h}}{2}%
(x^{2}+y^{2})\right]\psi(x,y) =E\psi(x,y).
\end{equation}

By using the same procedure as in the case of the free particle, we
diagonalize the Hamiltonian, thus obtaining
\begin{equation}
\widehat{H}=\hbar \Omega _{\alpha }\left( \widehat{\alpha }^{\dag }\widehat{%
\alpha }+\widehat{\beta }^{\dag }\widehat{\beta }+1\right) ,  \label{Hdab1}
\end{equation}%
where
\begin{eqnarray}
\Omega _{\alpha }&=&\omega _{h}+B_{h}=\sqrt{\frac{k}{m^{\ast }}+\frac{k_{0}}{%
m^{\ast }}\left( \frac{\varepsilon }{\varepsilon _{0}}\right) ^{2\alpha }}+
\notag \\
&&\frac{\eta _0}{2m\hbar}\left(\frac{\epsilon }{\epsilon _0}\right)^{\alpha}%
\left[1+\frac{km\theta _0}{\eta _0}\left(\frac{\epsilon }{\epsilon _0}%
\right)^{\beta -\alpha}\right],
\end{eqnarray}
where $\omega _{h}=\sqrt{K_{h}/m^{\ast }}$ is the generalized effective
frequency of the two-dimensional harmonic oscillator in the SQF
non-commutative algebra. The eigenvalues of $\widehat{H}$ can be written as
\begin{equation}
E=\hbar \Omega _{\alpha }\left( n_{\alpha }+n_{\beta }+1\right) ,
\label{Ev1nn}
\end{equation}%
where $n_{\alpha },n_{\beta }=0,1,\cdots $. The corresponding eigenstates
can be obtained as
\begin{equation}
|\psi _{n_{\alpha },n_{\beta }}\rangle =\frac{1}{\sqrt{n_{\alpha }!n_{\beta
}!}}\widehat{n}_{\alpha }\widehat{n}_{\beta }|0,0\rangle ,  \label{Wf1n}
\end{equation}%
where $\widehat{n}_{\alpha }=\widehat{\alpha }^{\dag }\widehat{\alpha }$ and
$\widehat{n}_{\beta }=\widehat{\beta }^{\dag }\widehat{\beta }$ are the
quasi - particle operators, and $|0,0\rangle $ is the ground state of the
two-dimensional harmonic oscillator in the SQF non-commutative algebra.

\section{Quantum dynamics in the Energy Coupling (EC) model}

\label{sect5}

In the present Section we investigate the two basic quantum mechanical
models, the free particle, and the harmonic oscillator, respectively, in the
Energy Coupling (EC) non-commutative algebra, by assuming that the
non-commutativity parameters $\theta $ and $\eta$ are functions of the
particle energy $E$ only, and independent of energy scale of the quantum
spacetime fluctuations.

\subsection{Quantum evolution - the Schr\"{o}dinger equation}

For the EC non-commutative algebra, the effective potential can be written
into a unified form for \textit{both the free particle and harmonic potential%
} as,
\begin{equation}  \label{V0n}
V_{\text{eff}} = -B_{h}(E)L_{z}+\frac{K_{h}(E)}{2}(x^{2}+y^{2}).
\end{equation}
Then the generalized Schr\"{o}dinger equation can be obtained as
\begin{equation}  \label{SchE1}
i\hbar \frac{\partial }{\partial t}\Psi(x,y,t)=\widehat{H}\Psi(x,y,t),
\end{equation}
where
\begin{equation}
\widehat{H} = -\frac{\hbar^{2}}{2m^{*}}\triangle_{2} -B_{h}(E)L_{z}+\frac{%
K_{h}(E)}{2}(x^{2}+y^{2}).
\end{equation}%
Since $\widehat{H}$ is independent of time, the wave function is of the form
\begin{equation}  \label{Ps2}
\Psi(t,x,y) =e^{-\frac{i}{\hbar }E t}\psi(x,y),
\end{equation}%
By substituting the wave function (\ref{Ps2}) into the Schr\"{o}dinger
equation (\ref{SchE1}), we obtain the stationary Schr\"{o}dinger equation as
\begin{eqnarray}  \label{ECSchE1}
\left[-\frac{\hbar^{2}}{2m^{*}}\triangle_{2} -B_{h}(E)L_{z}+\frac{K_{h}(E)}{2%
}(x^{2}+y^{2})\right]\psi= E\psi.  \notag \\
\end{eqnarray}

\subsection{Quantum evolution - wave function and energy levels}

In the following we obtain the solutions (wave functions) and the energy
levels of the Schr\"{o}dinger equation (\ref{ECSchE1}) in the EC
non-commutative algebra. In the polar coordinate system $\left( r,\phi
\right) $ with $x=r\cos \phi $, $y=r\sin \phi $, the angular momentum
operator is represented by $L_{z}=-i\hbar \frac{\partial}{\partial \phi}$.
Then the generalized stationary Schr\"{o}dinger equation can be expressed as
\begin{eqnarray}  \label{ECSchE2}
&&\Bigg[-\frac{\hbar^{2}}{2m^{*}}\left(\frac{\partial^{2} }{\partial r^{2}} +%
\frac{1}{r}\frac{\partial }{dr}+\frac{1}{r^{2}}\frac{\partial^{2}}{\partial
\phi^{2}}\right) +i\hbar B_{h}(E) \frac{\partial}{\partial \phi}+  \notag \\
&&\frac{K_{h}(E)}{2}r^{2}\Bigg] \psi(r,\phi)=E\psi(r,\phi).
\end{eqnarray}

Due to the axial symmetry of Eq. (\ref{ECSchE2}), the wave function can be
represented as
\begin{equation}
\psi (r,\phi )=R(r)e^{im_{\phi }\phi },  \label{ECPs1}
\end{equation}%
where $m_{\phi }=0,1,2,\cdots $. By substituting Eq. (\ref{ECPs1}) into Eq. (%
\ref{ECSchE2}), we obtain
\begin{eqnarray}
&&r^{2}R^{\prime \prime }(r)+rR^{\prime }(r)+\Bigg\{\frac{2m^{\ast }}{\hbar
^{2}}\left[ E+m_{\phi }\hbar B_{h}(E)\right] r^{2}-  \notag  \label{SchE3ii}
\\
&&m_{\phi }^{2}-\frac{m^{\ast }K_{h}(E)}{\hbar ^{2}}r^{4}\Bigg\}R(r)=0,
\end{eqnarray}%
where $R^{\prime }(r)=dR(r)/dr$. By introducing a new radial coordinate $\xi
$, defined as
\begin{equation}
\xi =\frac{\left[m^*K_h(E)\right]^{1/4}}{\sqrt{\hbar}}r,
\end{equation}%
and by denoting
\begin{equation}
C=\frac{2\sqrt{m^*}}{\hbar}\frac{E+m_{\phi}\hbar B_h(E)}{\sqrt{K_h(E)}},
\end{equation}%
Eq.~(\ref{SchE3ii}) takes the form
\begin{equation}  \label{Sche3iii}
\xi ^{2}\frac{d^{2}R\left( \xi \right) }{d\xi ^{2}}+\xi \frac{dR\left( \xi
\right) }{d\xi }+\left( C\xi ^{2}-\xi ^{4}-m_{\phi }^{2}\right) R\left( \xi
\right) =0,
\end{equation}
or, equivalently,
\begin{equation}  \label{108}
\xi \frac{d}{d\xi}\left(\xi \frac{dR(\xi)}{d\xi}\right)+\left( C\xi ^{2}-\xi
^{4}-m_{\phi }^{2}\right) R\left( \xi \right) =0.
\end{equation}

In the range of values of $\xi $ so that $C\xi ^{2}>>\xi ^{4}$, or,
equivalently,
\begin{equation}
\xi <<\frac{\left(4m^{\ast }\right)^{1/4}\left[ E+m_{\phi }\hbar B_{h}(E)%
\right]^{1/2} }{\hbar ^{1/2} K_{h}^{1/4}(E)},
\end{equation}%
Eq. (\ref{Sche3iii}) becomes
\begin{equation}
\xi ^{2}\frac{d^{2}R\left( \xi \right) }{d\xi ^{2}}+\xi \frac{dR\left( \xi
\right) }{d\xi }+\left(C \xi ^{2}-m_{\phi }^{2}\right) R\left( \xi \right)
=0,  \label{Sche3siii}
\end{equation}%
and has the general solution given by
\begin{equation}
R_{m_{\phi }}\left( \xi \right) =c_{1}J_{m_{\phi }}\left(\sqrt{C} \xi
\right) +c_{2}Y_{m_{\phi }}\left(\sqrt{C} \xi \right) ,  \label{sol3iii}
\end{equation}%
where $J_{m_{\phi }}\left( \xi \right) $ is the Bessel function of the first
kind, while $Y_{m_{\phi }}\left( \xi \right) $ denotes the Bessel function
of the second kind \cite{Stegun}. $c_{1}$ and $c_{2}$ denote two arbitrary
integration constants. Since the function $Y_{m_{\phi }}\left( \xi \right) $
is singular at the origin, we must take $c_{2}=0$ in the solution (\ref%
{sol3iii}). Therefore the general solution of Eq.~(\ref{Sche3siii}) can be
written as
\begin{eqnarray}
R_{m_{\phi }}\left( \xi \right) &=&c_{1}J_{m_{\phi }}\left(\sqrt{C} \xi
\right) =  \notag \\
&&c_{1}\sum_{l=0}^{\infty }\frac{\left( -1\right) ^{l}}{l!\Gamma \left(
l+m_{\phi }+1\right) }\left( \frac{\sqrt{C}\xi }{2}\right) ^{2l+m_{\phi }}=
\notag \\
&&\frac{c_{1}}{\pi }\int_{0}^{\pi }\cos \left( m_{\phi }\tau -\sqrt{C}\xi
\sin \tau \right) d\tau ,
\end{eqnarray}%
where $\Gamma \left( z\right) $ is the gamma function. For small values of
the argument, the wave function behaves like $R_{m_{\phi }}\left( \xi
\right) \approx \left[ 1/\Gamma \left( m_{\phi }+1\right) \right] \left(%
\sqrt{C} \xi /2\right) ^{m_{\phi }}$, while for large values of the argument
we have $R_{m_{\phi }}\left( \xi \right) \approx c_{1}\sqrt{2/\pi\sqrt{C}
\xi }\cos \left( \sqrt{C}\xi -m_{\phi }\pi /2-\pi /4\right) +e^{\left\vert
\mathrm{Im}\;(\xi )\right\vert }O\left( 1/\left\vert \xi \right\vert \right)
$, a relation valid for $\left\vert \arg \sqrt{C}\xi \right\vert <\pi $ \cite%
{Stegun}.

In the interval $\left[ 0,\xi _{s}\right] $, the Bessel functions satisfy
the condition $\int_{0}^{\xi _{s}}J_{m_{\phi }}\left( j_{m_{\phi }l}\frac{x}{%
\xi _{s}}\right) J_{m_{\phi }}\left( j_{m_{\phi }n}\frac{x}{\xi _{s}}\right)
xdx=\left( 1/2\right) \xi _{s}^{2}\left[ J_{m_{\phi }+1}\left( j_{m_{\phi
}l}\right) \right] ^{2}\delta _{ln }$, where $j_{m_{\phi }l}$ is the $l$th
zero of $J_{m_{\phi }}\left( \xi \right) $ \cite{Stegun}. There is a large
literature on the zeros of the Bessel functions, and for a review and some
recent results see \cite{Elbert}.

Now we consider the case in which in Eq.~(\ref{108}) the term $\xi ^{4}$
cannot be neglected. To solve Eq.~(\ref{108}) we introduce a new coordinate $%
\zeta$ defined as $\zeta =\xi ^2$. Then we obtain immediately $\xi \frac{d}{%
d\xi}=2\zeta \frac{d}{d\zeta}$, $\xi \frac{d}{d\xi}\left(\xi \frac{dR}{d\xi}%
\right)=4\left(\zeta \frac{d^2R}{d \zeta ^2}+\zeta \frac{dR}{d\zeta}\right)$%
, and Eq.~(\ref{108}) takes the form
\begin{equation}  \label{113}
\frac{d^2R(\zeta)}{d\zeta ^2}+\frac{1}{\zeta}\frac{dR(\zeta)}{d\zeta}+\left(%
\frac{C-\zeta}{4\zeta}-\frac{m_{\phi}^2}{4\zeta ^2}\right)=0.
\end{equation}
Next we introduce a new function $L\left(\zeta,m_{\phi}\right)$ by means of
the transformation
\begin{equation}
R\left(\zeta \right)=\zeta ^{m_{\phi}/2}e^{-\zeta/2}L\left(\zeta,
m_{\phi}\right).
\end{equation}
Hence Eq.~(\ref{113}) becomes
\begin{eqnarray}  \label{115}
&&\zeta \frac{d^2L\left(\zeta,m_{\phi}\right)}{d\zeta ^2}+\left(m_{\phi}+1-%
\zeta\right)\frac{dL\left(\zeta,m_{\phi}\right)}{d\zeta}+  \notag \\
&&\frac{1}{2}\left(\frac{C}{2}-m_{\phi}-1\right)L\left(\zeta,m_{\phi}%
\right)=0.
\end{eqnarray}
For
\begin{equation}
n=\frac{1}{2}\left(\frac{C}{2}-m_{\phi}-1\right)\geq 0, n\in \mathbf{\mathrm{%
N}},
\end{equation}
that is, for $n$ taking non-negative integer values, the well-behaved
solution of Eq.~(\ref{115}) is given by
\begin{equation}
L\left(\zeta,m_{\phi}\right)=cL_n^{\left(m_{\phi}\right)}\left(\zeta\right),
\end{equation}
where $c$ is an arbitrary integration constant, and $L_n^{\left(m_{\phi}%
\right)}\left(\zeta\right)$ are the generalized Laguerre polynomials,
defined as \cite{Stegun}
\begin{equation}
L_n^{\left(m_{\phi}\right)}\left(\zeta\right)=\frac{\zeta
^{m_{\phi}}e^{\zeta}}{n!}\frac{d^n}{d\zeta ^n}\left(e^{-\zeta}\zeta
^{n+m_{\phi}}\right),
\end{equation}
or, alternatively, as $L_n^{\left(m_{\phi}\right)}=x^{-m_{\phi}}\left(\frac{d%
}{dx}-1\right)^n\zeta ^{n+m_{\phi}}/n!$. Hence the physical solution of Eq.~(%
\ref{Sche3iii}) can be obtained as
\begin{equation}
R^{\left(m_{\phi }\right)}_n(\xi )=ce^{-\xi ^{2}/2}\xi ^{m_{\phi
}}L_{n}^{\left(m_{\phi }\right)}\left(\xi ^{2}\right) .  \label{wfp}
\end{equation}

The radial wave function must satisfy the normalization condition
\begin{equation}
\int_{0}^{\infty }{\left[R_{n}^{\left(m_{\phi}\right)}(r)\right]^{2}rdr}=1,
\end{equation}%
or, equivalently,
\begin{equation}
c\frac{\hbar}{\left[m^{*}K_h(E)\right]^{1/2}}\int_{0}^{\infty }\xi ^{2m_{\phi}}e^{-\xi ^{2}}\left[ L_{n}^{\left(m_{\phi}%
\right)}\left(\xi ^{2}\right) \right] ^{2}\xi d\xi =1.
\end{equation}

By introducing a new variable $\xi ^{2}=x$, $2\xi d\xi =dx$, we obtain
\begin{equation}
\frac{c}{2}\frac{\hbar}{\left[m^{*}K_h(E)\right]^{1/2}}\int_{0}^{\infty }{x^{m_{\phi}}e^{-x}\left[ L_{n}^{\left(m_{\phi}%
\right)}\left( x\right) \right] ^{2}dx}=1.
\end{equation}

By taking into account the mathematical identity \cite{Stegun, Landau}
\begin{equation}
\int_{0}^{\infty }e^{-x}x^{a}\left( L_{n}^{(a)}\left( x\right) \right)
^{2}dx=\left( n+a\right) !/n! ,
\end{equation}
we find for the integration constant $c$ the expression
\begin{equation}
c=\frac{2n!}{\left(n+m_{\phi}\right)!}\frac{\left[m^{*}K_h(E)\right]^{1/2}}{\hbar}.
\end{equation}

The quantized energy levels of the free particle, and of the harmonic
oscillator, can be obtained in the noncommutative energy dependent quantum
mechanics as solutions of the algebraic equation
\begin{equation}  \label{125}
\frac{\hbar}{\sqrt{m^*}}\left(2n+m_{\phi}+1\right)=\frac{E+m_{\phi}\hbar
B_h(E)}{\sqrt{K_h(E)}}.
\end{equation}

The commutative quantum mechanical limit for \textit{the harmonic oscillator}
is recovered, as one can see easily from Eqs.~(\ref{59a})-(\ref{59c}), when $E<<E_{0}$,
giving $B_{h}(E)=0$ and $K_{h}(E)=k$, respectively. Then from Eq.~(\ref{125}%
) we immediately obtain
\begin{equation}
E_{com}=E_{n}^{\left( m_{\phi }\right) }=\hbar \omega \left( 2n+m_{\phi }+1\right) ,
\end{equation}%
where $\omega ^{2}=k/m$. This relation gives the energy spectrum of the harmonic oscillator in commutative quantum mechanics \cite{clim}.

\subsection{The case of the free particle}

Let's consider first the simple case of the free particle, with $%
B_{e}(E)=\left. B_{h}(E)\right\vert _{k=0}=B_{0}\left( E/E_{0}\right)
^{\alpha }$, $k_{e}(E)=\left. K_{h}(E)\right\vert _{k=0}=\left(
k_{0}/2\right) \left( E/E_{0}\right) ^{2\alpha }$, and $m^{\ast }=m$,
respectively. Then from Eq.~(\ref{125}) it follows that the energy levels of
the free particle are given by
\begin{eqnarray}
E_{n}^{\left( m_{\phi }\right) } &=&\left( \sqrt{\frac{2m}{\hbar ^{2}k_{0}}}%
E_{0}\right) ^{1/(\alpha -1)}\times   \notag  \label{121} \\
&&\frac{E_{0}}{\left[ 2n+\left( 1-B_{0}\sqrt{2m/k_{0}}\right) m_{\phi }+1%
\right] ^{1/(\alpha -1)}},\alpha \neq 1.  \notag \\
&&
\end{eqnarray}

For $\alpha =1$ the energy spectrum of the particle is continuous, and the
the two quantum numbers $n$ and $m_{\phi}$ must satisfy the condition $%
E_0=\hbar\sqrt{k_0/2m}\left[2n+\left(1-B_0\sqrt{2m/k_0}\right)m_{\phi}+1%
\right]$.

The wave function of the two-dimensional free particle in energy dependent
noncommutative quantum mechanics is given by
\begin{eqnarray}
R_{n}^{\left( m_{\phi }\right) }(r)&=&c\left( \frac{mk_{0}}{2\hbar ^{2}}%
\right) ^{\frac{m_{\phi }}{4}}\left( \frac{E}{E_{0}}\right) ^{\frac{m_{\phi
}\alpha } {2}}e^{-\sqrt{\frac{mk_{0}}{8\hbar ^{2}}}\left( \frac{E}{E_{0}}%
\right) ^{\alpha }r^{2}}\times  \notag \\
&&r^{m_{\phi }}L_{n}^{\left( m_{\phi }\right) }\left[ \sqrt{\frac{mk_{0}}{%
2\hbar ^{2}}}\left( \frac{E}{E_{0}}\right) ^{\alpha }r^{2}\right] .
\end{eqnarray}

In commutative quantum mechanics the wave function of a freely moving
quantum particle with momentum $\vec{p}$ is given by $\Psi _{\vec{p}}=%
\mathrm{const.}\;e^{i\vec{p}\cdot\vec{r}/\hbar}$. If we introduce the wave
vector $\vec{K}$, defined by $\vec{K}=\vec{p}/\hbar$, then the wave function
of the free particle is given by $\Psi _{\vec{K}}=\mathrm{const.}\;e^{i\vec{K%
}\cdot\vec{r}}$. The energy spectrum of the particle is continuous, with $E=%
\vec{p}^2/2m=\hbar^2K^2/2m$. The evolution of the free particle in the
energy-dependent noncommutative quantum mechanics is qualitatively different
from the standard quantum mechanical case. The particle is not anymore
"free", but its dynamics is determined by the presence of the effective
potential generated by the non-commutative effects. Moreover, the energy
levels are quantized in terms of two quantum numbers $n$ and $m_{\phi}$.

The ground state of the free particle corresponds to the choice $n=m_{\phi
}=0$. Then the ground state energy is given by

\begin{equation}
E_{0}^{(0)}=\left( \frac{2m}{\hbar ^{2}k_{0}}\right) ^{\frac{1}{2\left(
\alpha -1\right) }}E_{0}^{\frac{\alpha }{\alpha -1}},\alpha \neq 1.
\end{equation}

The radial wave function of the ground state of the free particle in
energy-dependent quantum mechanics takes the form

\begin{equation}
R_{0}^{\left( 0\right) }(r)=ce^{-\sqrt{\frac{mk_{0}}{8\hbar ^{2}}}\left(
\frac{E}{E_{0}}\right) ^{\alpha }r^{2}}.
\end{equation}

A possibility to test the energy-dependent noncommutative quantum mechanics would be through the study of the collision and scattering processes. Collisions are characterized by the differential cross section $d\sigma /d\Omega$, defined as the ratio of the number $N$ of particles scattered into direction $\left(\theta, \phi\right)$ per unit time per unit solid angle, divided by incident flux $j$, $d\sigma /d\Omega=N/j$ \cite{1a, Landau}.  Usually one considers that the incident wave on the target corresponds to a free particle, and the scattering wave function is given by $\psi \left(\vec{r}\right)\sim e^{i\vec{K}\cdot \vec{r}}+f\left(\theta\right)e^{i\vec{K}\cdot \vec{r}}/r$. However, in energy dependent quantum mechanics the wave function of the free particle at infinity cannot be described anymore as a simple plane wave. Hence, at least in principle, energy-dependent noncommutative effects could be determined and studied experimentally through their effects on the scattering cross sections in very high energy particle collisions. A cross section dependent on the particle energies may be an indicator of the noncommutative quantum mechanical effects, and may provide an experimental method to detect the presence of the quantum space-time.

\subsection{The harmonic oscillator}

Next we consider the harmonic oscillator problem in the Energy Coupling model of the energy-dependent noncommutative quantum mechanics. By taking into account the explicit forms of the effective mass, effective magnetic field and effective elastic constant as given by Eqs.~(\ref{59a})-(\ref{59c}), it follows that the energy spectrum of the harmonic oscillator is obtained as a solution of the nonlinear algebraic equation given by
\begin{eqnarray}
&&\hbar \omega \sqrt{1+\frac{m^{2}\omega ^{2}\theta _{0}}{4\hbar ^{2}}%
x^{\beta }}\left( 2n+m_{\phi }+1\right) =  \nonumber \\
&&\frac{E_{0}x+\frac{m_{\phi }\omega ^{2}}{2}\left[ \frac{\eta _{0}}{k}%
x^{\alpha }+m\theta _{0}x^{\beta }\right] }{\sqrt{1+\frac{\eta _{0}^{2}}{%
8m^{2}\omega ^{2}\hbar ^{2}}x^{2\alpha }}},
\end{eqnarray}%
where we have denoted $x=E/E_{0}$, and $\omega =\sqrt{k/m}$. In order to
solve this equation we need to fix, from physical considerations, the
numerical values of the quantities $\alpha $ and $\beta $. For arbitrary
values of $\alpha $ and $\beta $ the energy levels can be obtained generally
only by using numerical methods. In the first order approximation we obtain

\begin{eqnarray}
\hspace{-1.3cm}&&\hbar \omega \left( 1+\frac{m^{2}\omega ^{2}\theta _{0}}{8\hbar ^{2}}%
x^{\beta }\right) \left( 2n+m_{\phi }+1\right) =  \nonumber \\
\hspace{-1.3cm}&&E_{0}x+\frac{m_{\phi }\omega ^{2}}{2}\left[ \frac{\eta _{0}}{k}x^{\alpha
}+m\theta _{0}x^{\beta }\right] -
\frac{E_{0}\eta _{0}^{2}}{16m^{2}\omega
^{2}\hbar ^{2}}x^{2\alpha +1},
\end{eqnarray}%
or, equivalently,
\begin{eqnarray}
&&\left( 1+\frac{m^{2}\omega ^{2}\theta _{0}}{8\hbar ^{2}}x^{\beta }\right)
E_{com}=E_{0}x+\frac{m_{\phi }\omega ^{2}}{2}\times \nonumber\\
&&\left[ \frac{\eta _{0}}{k}%
x^{\alpha }+m\theta _{0}x^{\beta }\right]
-\frac{E_{0}\eta _{0}^{2}}{16m^{2}\omega ^{2}\hbar ^{2}}x^{2\alpha +1},
\end{eqnarray}%
where $E_{com}$ denotes the energy levels of the harmonic oscillator in the
commutative formulation of quantum mechanics. In the simple case $\alpha
=\beta =1$, in the first approximation we obtain for the energy levels the
algebraic equation
\bea
\hspace{-.3cm}&&\left( 1+\frac{m^{2}\omega ^{2}\theta _{0}}{8\hbar ^{2}}x\right) E_{com}=%
\left[ E_{0}+\frac{m_{\phi }\omega ^{2}}{2}\left( \frac{\eta _{0}}{k}%
+m\theta _{0}\right) \right] x\nonumber\\
\hspace{-0.3cm}&&-\frac{E_{0}\eta _{0}^{2}}{16m^{2}\omega
^{2}\hbar ^{2}}x^{3}.
\eea

By neglecting the term $x^{3}$, we obtain the energy levels in the
energy-dependent noncommutative quantum mechanics in the first order
approximation as

\begin{equation}
E_{n}^{\left( m_{\phi }\right) }\approx \frac{E_{com}}{E_{0}+\frac{m_{\phi
}\omega ^{2}}{2}\left( \frac{\eta _{0}}{k}+m\theta _{0}\right) -\frac{%
m^{2}\omega ^{2}\theta _{0}}{8\hbar ^{2}}E_{com}}.
\end{equation}

The wave function of the ground state of the energy-dependent harmonic
oscillator in noncommutative quantum mechanics, corresponding to $n=m_{\phi
}=0$,  can be written as

\begin{equation}
R_{0}^{(0)}\left( r\right) \sim \exp \left\{ -\frac{m\omega }{%
\hbar }\left[ \frac{1+\frac{\eta _{0}^{2}}{8m^{2}\omega ^{2}\hbar ^{2}}%
\left( \frac{E}{E_{0}}\right) ^{2\alpha }}{1+\frac{m^{2}\omega ^{2}\theta
_{0}}{4\hbar ^{2}}\left( \frac{E}{E_{0}}\right) ^{\beta }}\right] ^{1/2}%
\frac{r^{2}}{2}\right\} .
\end{equation}

In the limit $E<<E_{0}$, we recover the standard commutative result for the eave function of the ground state of the quantum mechanical harmonic oscillator, $\psi \left( r\right)\sim \exp \left\{ -%
\frac{m\omega }{\hbar }\frac{r^{2}}{2}\right\} $ \cite{Landau,1a}.

The wave function of the ground state of the harmonic oscillator in the energy-dependent noncommutative quantum mechanics can be written in a form analogous to the commutative case by introducing the effective energy-depending frequency $\omega _{eff}$, defined as
\begin{equation}
\omega _{eff}(E)=\omega \left[ \frac{1+\frac{\eta _{0}^{2}}{8m^{2}\omega
^{2}\hbar ^{2}}\left( \frac{E}{E_{0}}\right) ^{2\alpha }}{1+\frac{%
m^{2}\omega ^{2}\theta _{0}}{4\hbar ^{2}}\left( \frac{E}{E_{0}}\right)
^{\beta }}\right] ^{1/2}.
\end{equation}

Then the wave function of the ground state of the harmonic oscillator can be
written as
\begin{equation}
R_{0}^{(0)}\left( r\right) \sim\exp \left[ -\frac{m\omega _{eff}(E)%
}{\hbar }\frac{r^{2}}{2}\right] .
\end{equation}

Hence we have completely solved the problem of the quantum mechanical motion
of the free particle, and of a particle in a harmonic potential,
in the noncommutative quantum mechanics with energy-dependent strengths.

\section{Quantum evolution in the Energy Operator (EO) energy dependent
non-commutative geometry}\label{sect6}

Finally, we will consider in detail the third possibility of constructing
quantum mechanics in the framework of energy-dependent non-commutative
geometry. This approach consists \textit{in mapping the energy in the
non-commutative algebra to an operator}. As we have already discussed, we
have two possibilities to develop such an approach, by mapping the energy to
the time operator, or to the particle Hamiltonian. Under the assumption of a
power-law dependence on energy of the noncommutative strengths, \textit{in
the general case these maps lead to a fractional Schr\"{o}dinger equation}.
In the following we will first write down the basic fractional Schr\"{o}%
dinger equations for the free particle and the harmonic oscillator case, and
after that we will proceed to a detailed study of their properties. We will
concentrate on the models obtained by the time operator representation of
the energy. But before proceeding to discuss the physical implications of
the generalized Schr\"{o}dinger equation with fractional operators, we will
present o very brief summary of the basic properties of the fractional
calculus.

\subsection{Fractional calculus - a brief review}

For $\alpha ,\beta \notin N$, the quantum mechanical model introduced in the
previous Sections, based on the power law energy dependence of the
noncommutative strengths, leads to the interesting question of the
mathematical and physical interpretation of the operators of the form $\left[
\left( ih/E_{0}\right) \partial /\partial t\right] ^{\alpha }$. It turns out
that such operators can be written is terms of a \textit{fractional
derivative} as 
\begin{equation}
\eta \left( i\hbar \frac{\partial }{\partial t}\right) \Psi =\eta _{0}\left(
\frac{i\hbar }{E_{0}}\right) ^{\alpha }\frac{\partial ^{\alpha }}{\partial
t^{\alpha }}=\eta _{0}\left( \frac{i\hbar }{E_{0}}\right) ^{\alpha
}D_{t}^{\alpha }\Psi ,
\end{equation}
where $D_{t}^{\alpha }=\partial ^{\alpha }/\partial t^{\alpha }$ is the
fractional derivative of $\Psi $ of the order $\alpha >0$, which can be
defined in terms of the fractional integral $D^{-m}_t\Psi (t)$ as \cite%
{fract1,fract2,Hermann}
\begin{equation}
D_{t}^{\alpha }\Psi \left( t\right) =D_{t}^{m}\left[ D_{t}^{-\left( m-\alpha
\right) }\Psi \left( t\right) \right] .
\end{equation}%
Hence in fractional calculus a fractional derivative is defined via a
fractional integral. By interpreting the time derivative operators as
fractional derivatives we obtain the fractional Schr\"{o}dinger equations as
given by the equations presented in the next Section.

There are several definitions of the fractional derivative that have been
intensively investigated in the mathematical and physical literature. For
example, the Caputo fractional derivative is defined as
\begin{equation}
_{\alpha }^{C}D_{t}^{\alpha }\Psi \left( t\right) =\frac{1}{\Gamma \left(
n-\alpha \right) }\int_{a}^{t}\frac{\Psi ^{(n)}(\tau )d\tau }{\left( t-\tau
\right) ^{\alpha +1-n}}.
\end{equation}

Given a function $f(x)=\sum_{k=0}^{\infty }a_{k}x^{k\alpha }$, its
fractional Caputo derivative can be obtained according to
\begin{equation}
\frac{d^{\alpha }}{dx^{\alpha }}f(x)=\sum_{k=0}^{\infty }a_{k+1}\frac{\Gamma %
\left[ 1+(k+1)\alpha \right] }{\Gamma \left( 1+k\alpha \right) }x^{k\alpha },
\end{equation}%
where $\Gamma \left( z\right) $ is the gamma function defined as $\Gamma
\left( z\right) =\int_{0}^{\infty }t^{z-1}e^{-t}dt$, with the property $%
\Gamma \left( 1+z\right) =z\Gamma \left( z\right) $ \cite{Hermann}.

Consequently, we obtain the definition of the Caputo fractional derivative
of the exponential function as
\begin{eqnarray}
\frac{d^{\alpha }}{dx^{\alpha }}e^{x}&=&\frac{d^{\alpha }}{dx^{\alpha }}%
\sum_{n=0}^{\infty }\frac{x^{n}}{n!}=\sum_{n=1}^{\infty }\frac{x^{n-\alpha }%
}{\Gamma \left( 1+n-\alpha \right) }=  \notag \\
&&x^{1-\alpha }E_{1,2-\alpha }\left( x\right) ,
\end{eqnarray}%
where $E_{1,2-\alpha }(x)$ is the generalized Mittag-Leffler function,
defined as $E_{\alpha ,\beta }\left( z\right) =\sum_{n=0}^{\infty
}z^{n}/\Gamma \left( n\alpha +\beta \right) $ \cite{Hermann}.

Another definition of the fractional derivative is the Liouville definition,
which implies \cite{Hermann}
\begin{equation}
\frac{d^{\alpha }}{dx^{\alpha }}e^{kx}=k^{\alpha }e^{kx},k\geq 0.
\end{equation}
Finally, we also point out the definition of the left-sided
Riemann-Liouville fractional integral of order $\nu$ of the function $f(t)$,
which is defined as
\begin{equation}
\hspace{-0.1cm}_aD^{\nu}f(x):=\frac{1}{\Gamma (n-\nu)}\frac{d^n}{dx^n}\left(\int_0^x{\frac{%
f(\tau}{x-\tau}^{\nu+1-n}d\tau }\right),
\end{equation}
a definition which is valid for $n-1<\nu <n\in \mathbb{N}$ \cite{Hermann}.

\subsection{The fractional Schr\"{o}dinger equation}

In the present Section we will consider the evolution of a system in the Energy Operator representation of the energy-dependent noncommutative quantum mechanics. We will restrict again our analysis to the two-dimensional
case only. The generalized
fractional two-dimensional Schr\"{o}dinger equation can then be expressed as
\begin{equation}  \label{SchE-o}
i\hbar \frac{\partial }{\partial t}\Psi(x,y,t)=\widehat{H}\Psi(x,y,t),
\end{equation}
where
\begin{equation}  \label{FH}
\widehat{H}= -\frac{\hbar^{2}}{2m_{\ell}^{*}}\triangle_{2}+V_{\text{eff}},
\end{equation}
with
\begin{equation}
V_{\text{eff}} =V(x,y) -B_{\ell}D_{\ell}^{\alpha}L_{z} + \frac{1}{2}%
K_{\ell}D_{\ell}^{2\alpha}(x^{2}+y^{2}),
\end{equation}%
and we have denoted
\begin{subequations}
\label{Lda3}
\begin{eqnarray}
\frac{1}{m_{\ell}^{*}} &=& \frac{1}{m}+\frac{2\kappa_{\ell}}{\hbar^{2}}%
,\quad \kappa_{\ell} = \frac{k\theta _{\alpha}^{2}}{8}, \\
B_{\ell} &=& B_{\alpha}^{\ell} + \frac{k\theta _{\alpha}}{2\hbar }, \quad
K_{\ell}= k+k^{\alpha}_{\ell},
\end{eqnarray}
where $\ell=I,II$ denote the two different operator representations of the
EO non-commutative algebra. $D_{I}^{\alpha}\equiv \left(\frac{\partial}{%
\partial t}\right)^{\alpha} $ and $D_{II}^{\alpha}\equiv
D_{x}^{2\alpha}+D_{y}^{2\alpha}$ are the fractional derivatives with respect
to time and space when $\alpha$ is a rational integer.

In the following we will concentrate only on the time operator
representation of the non-commutative quantum mechanics with energy
dependent strengths, namely $\ell=I$. Thus, by representing the wave
function as

\end{subequations}
\begin{equation}
\Psi \left( t,x,y\right) =e^{-\frac{i}{\hbar }E t}\psi \left( x,y\right) ,
\end{equation}%
the Schr\"{o}dinger equation becomes a fractional differential equation
given by 
\begin{eqnarray}  \label{EOSchE-3}
&&\Bigg[ -\frac{\hbar^{2}}{2m_{\ell}^{*}}\triangle_{2}+V(x,y) -B_{\ell}\mathcal{D}%
_{\alpha }(t)L_{z} +  \notag \\
&&\frac{1}{2}K_{\ell}\mathcal{D}_{2\alpha }(t) (x^{2}+y^{2})\Bigg] \psi
\left( x,y\right) = E\psi(x,y),
\end{eqnarray}
where
\begin{equation}
\mathcal{D}_{\alpha }(t)=e^{\frac{i}{\hbar }E t}D_{t}^{\alpha }e^{-\frac{i}{%
\hbar }E t}, \quad \mathcal{D}_{2\alpha }(t)=e^{\frac{i}{\hbar }E
t}D_{t}^{2\alpha }e^{-\frac{i}{\hbar }E t}.
\end{equation}

If the function $e^{-\frac{i}{\hbar }E t}$ is an eigenfunction of the
fractional derivation operators $D^{\alpha }$ and $D^{2\alpha }$, so that
\begin{equation}
D_{t}^{\alpha }e^{-\frac{i}{\hbar }E t}=a_{\alpha }e^{-\frac{i}{\hbar }E t},
\quad D_{t}^{2\alpha }e^{-\frac{i}{\hbar }E t}=a_{2\alpha }e^{-\frac{i}{%
\hbar }E t},
\end{equation}%
where $a_{\alpha }$ and $a_{2\alpha }$ are constants, the separation of the
time variable can be performed in the noncommutative fractional Schr\"{o}%
dinger equation with energy dependent noncommutative strengths.

\subsection{The free particle-the case $\alpha =1$}

For simplicity, in the following we investigate the quantum dynamics in the Energy Operator representation only in the case of the free
particle, by assuming $V(\widehat{x},\widehat{y})=0$. Therefore $k=0$, and the effective mass of
the particle coincides with the ordinary mass, $m_l^*=m$. Moreover, for
simplicity we will restrict our analysis to the choice $\alpha =1$. Then $%
\mathcal{D}_{1}(t)=-\frac{i}{\hbar }E$ and $\mathcal{D}_{2}(t)=-\frac{1}{%
\hbar ^{2}}E^{2}$, respectively.

Explicitly, the Schr\"{o}dinger equation describing the motion of the free
particle in the energy-dependent noncommutative geometry takes the form
\begin{eqnarray}
\hspace{-1.2cm} &&i\hbar \frac{\partial }{\partial t}\Psi (t,x,y)-\frac{%
i\eta _{0}}{2mE_{0}}\hat{L}_{z}\frac{\partial }{\partial t}\Psi (t,x,y)+
\notag \\
\hspace{-1.2cm} &&{\frac{\eta _{0}^{2}}{8mE_{0}^{2}}\left(
x^{2}+y^{2}\right) }\frac{\partial ^{2}}{\partial t^{2}}\Psi (t,x,y)=-\frac{%
\hbar ^{2}}{2m}\Delta _{2}\Psi (t,x,y).
\end{eqnarray}

In the polar coordinate system $\left( r,\phi \right) $ with $r=r\cos \phi $%
, $y=r\sin \phi $, we have%
\begin{equation}
\hat{L}_{z}=\frac{\hbar }{i}\frac{\partial }{\partial \phi },
\end{equation}%
and
\begin{equation}
\Delta _{2}=\frac{1}{r}\frac{\partial }{\partial r}\left( r\frac{\partial }{%
dr}\right) +\frac{1}{r^{2}}\frac{\partial ^{2}}{\partial \phi ^{2}},
\end{equation}%
respectively.

By introducing for the wave function the representation $\Psi
(t,x,y)=e^{-\left( i/\hbar \right) Et}\psi \left( r,\phi \right) $, and,
similarly to the previous Section, representing the reduced wave function as
$\psi \left( r,\phi \right) =R(r)e^{im_{\phi }\phi },m_{\phi }\in \mathbb{N}$%
, it follows that Eq.~(\ref{EOSchE-3}) takes the form
\begin{eqnarray}
&&r^{2}R^{\prime \prime }(r)+rR^{\prime }(r)+\Bigg[\frac{2m}{\hbar ^{2}}%
E\left( 1+m_{\phi }B_{I}\right) r^{2}-  \notag  \label{Scheiv} \\
&&\frac{mK_{I}}{\hbar ^{4}}E^{2}r^{4}-m_{\phi }^{2}\Bigg]R(r)=0.
\end{eqnarray}

By introducing a new independent variable $\xi $, defined as
\begin{equation}
r=\frac{\hbar}{\left(mK_IE^2\right)^{1/4}}\xi,
\end{equation}%
and by denoting
\begin{equation}
\sigma=\frac{2\sqrt{m}\left(1+m_{\phi}B_I\right)}{\sqrt{K_{I}}},
\end{equation}%
Eq.~(\ref{Scheiv}) takes the form
\begin{equation}
\xi ^{2}\frac{d^{2}R}{d\xi ^{2}}+\xi \frac{dR}{d\xi }+\left(\sigma  \xi ^{2}-\xi
^{4}-m_{\phi }^{2}\right) R=0,  \label{EOSchE-6}
\end{equation}%
When $\sigma \xi ^{2}>>\xi ^{4}$, or, equivalently, $\xi <<\left(4m\right)^{1/4}\left(
1+m_{\phi }B_{I}\right)^{1/2} /\left(K_{I}\right)^{1/4}$, that is, when the noncommutative
effects can be neglected, in the standard quantum mechanical limit we obtain
the solution of Eq. (\ref{EOSchE-6}) as
\begin{equation}
R(\xi )=c_{1}J_{n}(\xi )+c_{2}Y_{n}(\xi ),
\end{equation}%
where $c_{1}$ and $c_{2}$ are arbitrary integration constants, and $%
J_{n}(\xi )$ and $Y_{n}(\xi )$ are the Bessel function of the first kind and
the Bessel function of the second kind \cite{Stegun}, respectively. We have
already discussed in detail in the previous Section the behavior of the wave
function in this case.

If the term $\xi ^{4}$ cannot be neglected as compared to $\sigma \xi ^{2} $, then
the general solution of Eq. (\ref{EOSchE-6}) is given by
\begin{eqnarray}
R^{\left(m_{\phi }\right)}_n(\xi )=ce^{-\xi ^{2}/2}\xi ^{m_{\phi
}}L_{n}^{\left(m_{\phi }\right)}\left(\xi ^{2}\right) ,
\end{eqnarray}
that is, the same form of the solution as the one already considered when
discussing the evolution of the free particle and of the harmonic oscillator
in the Energy Coupling model of the energy-dependent non-commutative quantum
mechanics. The wave function must be finite at the origin. The normalization and other properties of the wave function
are similar to  the ones already investigated in the previous Section.

\section{Discussions and final remarks}\label{sect7}

In the present paper we have considered the quantum mechanical implications
of a noncommutative geometric model in which the strengths of the
noncommutative parameters are energy-dependent. From a physical point of
view such an approach may be justified since the effects of the
noncommutativity of the spacetime are expected to become apparent at
extremely high energies, of the order of the Planck energy, and at distance
scales of the order of the Planck length. By assuming an energy dependent
noncommutativity we obtain a smooth transition between the maximally
noncommutative geometry at the Planck scale, and its commutative ordinary
quantum mechanical version, which can be interpreted as the low energy limit
of the noncommutative high energy quantum mechanics. Hence this approach
unifies in a single formalism two apparently distinct approaches, the
noncommutative and commutative versions of quantum mechanics, respectively,
and generally leads to an energy dependent Schr\"{o}dinger equation, as
already considered in the literature \cite%
{End1,End2,End3,End3a,End4,End5,End6}.

One of the important question related to the formalism developed in the
present paper is related to the physical implications of the obtained
results. In the standard approach to noncommutative geometry, by using the
linearity of the $D$ map, one could find a representation of the
noncommutative observables as operators acting on the conventional Hilbert
space of ordinary quantum mechanics. More exactly, the $D$ map converts the
noncommutative system into a modified commutative quantum mechanical system
that contains an explicit dependence of the Hamiltonian on the
noncommutative parameters, and on the particular $D$ map used to obtain the
representation. The states of the considered quantum system are then wave
functions in the ordinary Hilbert space, the dynamics is determined by the
standard Schr\"{o}dinger equation with a modified Hamiltonian that depends
on the noncommutative strengths $\theta $ and $\eta$ \cite{17a}. Even that
the mathematical formalism is dependent on the functional form of the
adopted $D$ map that is used to realize the noncommutative-commutative
conversion, this is not the case for physical predictions of the theory such
as expectation values and probability distributions \cite{17a}. On the other
hand it is important to point the fact that the standard formalism in which
the energy dependence is ignored is not manifestly invariant under a
modification of the $D$ map.

In the energy-dependent approach to noncommutative geometry after performing
the $D$ map we arrive at an energy dependent Schr\"{o}dinger equation, which
contains some energy dependent potentials $V\left(x,E,\theta, \eta\right)$.
In order to obtain a consistent physical interpretation we need to redefine
the probability density, the normalization condition and the expectation
values of the physical observables. For example, in order to be sure that a
solution of the Schr\"{o}dinger equation associated with a stationary energy
$E$ is normalizable the following two conditions must hold simultaneously
\cite{End2},
\begin{equation}
1-\frac{\partial V\left(x,E,\theta, \eta\right)}{\partial E}\geq 0, x\in D,
N(\psi)<\infty.
\end{equation}

Moreover, in opposition to the standard case of an energy-independent
potential, the nonnegativity and the existence of the norm integral must
hold at the same time. The modified forms of the probability density and of
the probability current also lead to adjustments in the scalar product and
the norm that do not appear in standard quantum mechanics. Similarly to the
case of standard noncommutative quantum mechanics we also expect that,
similarly to the energy-independent case, our present formalism is not
independent under a change in the functional form of the $D$ map. The
phase-space formulation of a noncommutative extension of quantum mechanics
in arbitrary dimension, with both spatial and momentum noncommutativities,
was considered in \cite{17a}. By considering a covariant generalization of
the Weyl-Wigner transform and of the Darboux $D$ map, an
isomorphism between the operator and the phase-space representations of the
extended Heisenberg algebra was constructed. The map allows to develop a
systematic approach to derive the entire structure of noncommutative quantum
mechanics in phase space. More importantly, it turns out that the entire
formalism is independent of the particular choice of the Darboux map. The
extension of the results of \cite{17a} to the energy-dependent case would
help to clarify the mathematical structure and physical properties of
energy-dependent noncommutative quantum mechanics.

In order to implement the idea of energy-dependent noncommutativity we need
to specify the relevant energy scales. In the present work we have assumed a
two scale and a single energy scale model. Moreover, we have limited our
investigations to the case in which the noncommutative strengths have a
simple power law dependence on the energy. In the first approach the energy
dependence of the noncommutative strengths is determined by a specific
energy scale, which is related to the energy of the quantum fluctuations
that modify the geometry. This approach may be valid to describe physics
very nearby the Planck scale, where the vacuum energy may be the dominant
physical effect influencing the quantum evolution of particles in the
noncommutative geometric setting. In this context we have considered the
dynamics of two simple but important quantum systems, the free particle, and
the harmonic oscillator, respectively. The physical characteristics of the
evolution is strongly dependent on the energy of the quantum fluctuations,
with the oscillations frequencies effectively determined by the spacetime
fluctuations scale.

In our second model we have assumed that all the noncommutative effects can
be described by means of the particle energy scale, which is the unique
scale determining the physical implications of noncommutative geometry. The
choice of a single energy scale allows the smooth transition from the
noncommutative algebra of the Planck length to the commutative Heisenberg
algebra of the ordinary quantum mechanics that gives an excellent
description of the physical processes on the length and energy scales of the
atoms and molecules, and for the standard model of elementary particles. In
this case the basic physical parameters of the of the quantum dynamics of
the free particles and of the harmonic oscillator are energy dependent, with
the oscillation frequencies described by complicated functions of the
particle energy. Such an energy dependence of the basic physical parameters
of the quantum processes may have a significant impact on the high energy
evolution of the quantum particles.

Perhaps the most interesting physical implications are obtained in the
framework of our third approach, which consists in mapping the energy with a
quantum operator. There are two such possibilities we have briefly
discussed, namely, mapping the energy with the time derivative operator, and
with the Hamiltonian of the free particle (its kinetic energy). The
corresponding Schr\"{o}dinger equation changes its mathematical form,
content an interpretation, becoming a fractional differential equation, in
which the ordinary derivatives of quantum mechanics are substituted by
fractional ones.

Fractional Schr\"{o}dinger equations have been introduced for some time in
the physical literature \cite{fSch1}, and presently they are becoming a very
active field of research in both physics and mathematics \cite%
{fSch1,fSch2,fSch3,fSch4,fSch5,fSch6,fSch7,fSch8,fSch9, fSch9a,
fSch10,fSch11,fn1,fn2,fn3,fn4,fn5,fn6,fn7,fn8}. For an extensive review of fractional quantum mechanics see \cite{Laskin}. A typical example of a
fractional Schr\"{o}dinger equation is given by the equation \cite{fSch1}
\begin{equation}
i\hbar\frac{\partial \psi \left(\vec{r},t\right)}{\partial t}%
=D_{\alpha}\left(-\hbar ^2\Delta\right)^{\alpha /2}\psi \left(\vec{r}%
,t\right)+V \left(\vec{r},t\right)\psi \left(\vec{r},t\right),
\end{equation}
where $D_{\alpha}$ is a constant, and $\alpha $ is an arbitrary number. The
fractional Hamilton operator is Hermitic, and a parity conservation law for
fractional quantum mechanics can also be established. The energy spectra of
a hydrogenlike atom can also be obtained, while in this approach the
fractional oscillator with the Hamiltonian
\begin{equation}
H_{\alpha ,\beta }=D_\alpha |p|^\alpha +q^2|x|^\beta,
\end{equation}
where $\alpha $, $\beta $, and $q$ are constants, has the energy levels
quantized according to the rule \cite{fSch1}
\begin{equation}
E_n=\left[ \frac{\pi \hbar \beta D_\alpha ^{1/\alpha }q^{2/\beta }}{%
2B\left(\frac 1\beta ,\frac 1\alpha +1\right)}\right] ^{\frac{\alpha \beta }{%
\alpha +\beta }} \left(n+\frac 12\right)^{\frac{\alpha \beta }{\alpha +\beta
}},
\end{equation}
where the $B(a,b)$ function is defined by the integral representation $%
B(a,b)=\int\limits_0^1duu^{a-1}(1-u)^{b-1}$ \cite{Stegun}, and $\alpha $ and
$\beta$ are arbitrary numerical parameters. As for the physical origin of
the fractional derivatives, in \cite{La1,La2} it was shown that it
originates from the path integral approach to quantum mechanics. More
exactly, the path integral over Brownian trajectories gives the standard Schr%
\"{o}dinger equation of quantum mechanics, while the path integral over L$%
\acute{\mathrm{e}}$vy trajectories generates the fractional Schr\"{o}dinger
equation. In the present paper we have outlined the possibility of {\it another
physical path towards the fractional Schr\"{o}dinger equation}, namely, the
framework {\it of the quantum operator approach to energy dependent
noncommutative geometry}.

An interesting theoretical question in the field of noncommutative quantum mechanics  is the problem of the nonlocality generated by the  dynamical noncommutativity. This problem was investigated in \cite{34} for noncommutative quantum system with the coordinates satisfying the commutation relations $\left[\widehat{q}^i,\widehat{q}^j\right]=i\theta f(\sigma)\epsilon ^{ij}$, where $f(\sigma)$ is a function of the physical parameter $\sigma $, that could represent, for example, position, spin, energy etc. Then for the uncertainty relation between the coordinate operators $\widehat{x}$ and $\widehat{y}$ we obtain the uncertainty relation $\left(\Delta \widehat{x}\right)_{\Psi}\left(\Delta \widehat{x}\right)_{\Psi}\geq \left(\theta /2\right)\left|\left<\Psi\left|f(\sigma)\right|\Psi\right>\right|$ \cite{34}. As was pointed out in \cite{34},  if $|\Psi>$ represents a stationary state of the quantum system, that is, an eigenstate of the Hamiltonian, it follows that {\it the nonlocality induced by the noncommutativity of the coordinates will be a function of the energy}. Moreover, in the present approach to noncommutative quantum mechanics, the noncommutative strength $\theta$ is an explicit function of the energy, and therefore an explicit dependence of the nonlocality on the energy always appears. As a particular case we consider that $f(\sigma)=1$, that is the noncommutative strengths depend only on the energy. Then, by taking into account the normalization of the wave function we obtain  $\left(\Delta \widehat{x}\right)_{\Psi}\left(\Delta \widehat{x}\right)_{\Psi}\geq \left(\theta (E)/2\right)$, or by considering the explicit choice of the energy dependence of $\theta$ adopted in the present study, $\left(\Delta \widehat{x}\right)_{\Psi}\left(\Delta \widehat{x}\right)_{\Psi}\geq \left(\theta_0 /2\right)\left(E/E_0\right)^{\beta}$. Hence when $E<<E_0$, $\left(\Delta \widehat{x}\right)_{\Psi}\left(\Delta \widehat{x}\right)_{\Psi}\approx 0$, and we recover the standard quantum mechanical result. In the case of the harmonic oscillator the uncertainty relations for  the noncommutative coordinates can be obtained for $f(\sigma)=\sigma$ as $\left(\Delta \widehat{x}\right)_{\Psi}\left(\Delta \widehat{x}\right)_{\Psi}\geq O\left(\theta ^4\right)$, that is, nonlocality does not appear in higher orders of $\theta$ \cite{34}. For the case $f(\sigma)=\sigma ^2$, one finds $\left(\Delta \widehat{x}\right)_{\Psi}\left(\Delta \widehat{x}\right)_{\Psi}\geq \left(\theta /2\omega _{\sigma}\right)\left(n+1/2\right)+O\left(\theta ^3\right)$, where $\omega _{\eta}$ is the oscillation frequency of the $\sigma$ - dependent potential term in the total Hamiltonian, given by $V\left(\sigma\right)=\omega _{\sigma}^2 \sigma ^2/2$  \cite{34}.

A central question in the noncommutative extensions of quantum mechanics is the likelihood of its observational or experimental testing. A possibility of detecting the existence of the noncommutative phase space by using the Aharonov-Bohm effect was suggested in \cite{89}. As we have already seen the noncommutativity  of the momenta leads to the generation of an effective magnetic field and of an effective flux. In a mesoscopic ring this flux induces a persistent current. By using this effect it may be possible to detect the effective
magnetic flux generated by the presence of the noncommutative phase space, even it is very weak.
Persistent currents and magnetic fluxes in mesoscopic rings can be studied by using experimental methods developed in
nanotechnology \cite{89}.  The dynamics of a free electron in the two dimensional noncommutative phase space is equivalent to the evolution of the electron in an effective magnetic field,  induced by the effects of the noncommutativity of the coordinates and momenta.

For the motion of a free electron in the noncommutative phase space, the Hamiltonian can
be obtained as
\begin{equation}
H_{nc}=\frac{1}{2m}\left( \widehat{p}_{x}^{2}+\widehat{p}_{y}^{2}\right) =%
\frac{1}{2m^{\ast }}\left[ \left( p_{x}+eA_{x}\right) ^{2}+\left(
p_{y}+eA_{y}\right) ^{2}\right],
\end{equation}
where we have denoted by $m^{\ast }=m/\alpha $\ the effective mass in the noncommutative phase
space. The parameter $\alpha$ is defined through the relation $\theta \eta =2\hbar ^2\alpha ^2\left(1-\alpha ^2\right)$.  The components $A_x$ and $A_y$ of the effective vector potential $\vec{A}$ are given by
$A_{x}=\left(\eta /2e\alpha ^{2}\hbar \right)y$ and $A_{y}=-\left(\eta /2e\alpha ^{2}\hbar \right)x$, respectively \cite{89}, while the
effective magnetic field is obtained in the form $B_{z}=\eta/e\alpha ^{2}\hbar $.

A possibility of experimentally implementing a method that could detect noncommutative quantum mechanical effects consists in considering a one-dimensional ring in an external magnetic field $\vec{B}$, oriented  along the
axis of the ring. $\vec{B}$ is constant inside $r_{c}<R$ (ring
radius), which implies that the electrons are located only in the field-free region of the small ring. Moreover,  the
quantum electronic states are functions of the total magnetic flux crossing the ring only. By introducing a
polar coordinate system by means of the definitions $x=R\cos \varphi$, $y=R\sin \varphi$, we obtain for the
Hamiltonian of the electrons the expression \cite{89}
\begin{equation}
H_{nc}=-\frac{\hbar ^{2}}{2m^{\ast }R^{2}}\left[ \frac{\partial }{\partial
\varphi }+i\left( \frac{\phi }{\phi _{0}}-\frac{\phi _{nc}}{\phi _{0}}%
\right) \right] ^{2}-\frac{3\hbar ^{2}}{8m^{\ast }R^{2}}\frac{\phi _{nc}^{2}%
}{\phi _{0}^{2}},
\end{equation}
where $\phi _{nc}=2\pi R^{2}\eta/ e\hbar \alpha ^{2}$
represents {\it an effective magnetic flux} coming from the noncommutative phase space, while
$\phi _{0}=h/e$ is the quantum of the magnetic flux. By $\phi$ we have denoted the external
magnetic flux in the ring. Hence {\it noncommutative effects generate a persistent current in the ring},  which depends on the external and the
effective magnetic fluxes, respectively. The relation between the persistent current and the magnetic flux may provide a method to detect the existence of noncommutative quantum mechanical effects. Hence by considering a mesoscopic ring system in the presence of an external magnetic field, and by studying the relation between the persistent current and the external magnetic flux $\varphi$ one can infer the possible existence of noncommutative quantum mechanical effects \cite{89}. The theoretical model behind this experimental procedure can be easily reformulated by taking into account the variation of $\eta $ with the energy of the electrons. Hence the study of the persistent currents in mesoscopic systems by using experimental techniques already existent in nanotechnology may open the possibility of proving the existence of  new quantum mechanical physical structures that becomes dominant at high particle energies.

The investigation of the spacetime structure and physical processes at very
high energies, and small microscopic length scales may open the possibility
of a deeper understanding of the nature of the fundamental interactions and
of their mathematical description. In the present work we have developed
some basic tools that could help to give some new insights into the complex
problem of the nature of the quantum dynamical evolution processes at
different energy scales, and of their physical implications.

\section*{Acknowledgments}

We would like to thank the two anonymous reviewers for their comments and
suggestions that helped us to significantly improve our manuscript. T. H.
would like to thank the Yat Sen School of the Sun Yat Sen University in
Guangzhou, P. R. China, for the kind hospitality offered during the
preparation of this work. S.-D. L. thanks the Natural Science Foundation of Guangdong Province for financial support (grant No. 2016A030313313).


\begin{thebibliography}{99}
\bibitem{1} S. Doplicher, K. Fredenhagen, and J. E. Roberts, Commun. Math.
Phys. \textbf{172}, 187 (1995).

\bibitem{Landau} A. Messiah, Quantum Mechanics, Dover Publications, New
York, 1999

\bibitem{1a} L. D. Landau and E. M. Lifshitz, Quantum Mechanics:
Non-Relativistic Theory, Butterworth-Heinemann, Oxford, 2003

\bibitem{1b} S.-D. Liang, Quantum Tunneling and Field Emission Theories,
World Scientific, New Jersey, London, Singapore, 2014

\bibitem{2} K. Fredenhagen, Reviews in Mathematical Physics \textbf{7}, 559
(1995).

\bibitem{3} H. S. Snyder, Phys. Rev. \textbf{71}, 38 (1947).

\bibitem{4} C. N. Yang, Phys. Rev. \textbf{72}, 874, (1947).

\bibitem{5} A. Connes, Inst. Hautes \'{E}tudes Sci. Publ. Math. \textbf{62},
257 (1985).

\bibitem{6} V. G. Drinfel'd, Proc. of the International Congress of
Mathematicians (Berkeley, 1986), American Mathematical Society (1987).

\bibitem{7} S. L. Woronowicz, Publ. Res. Inst. Math. Sci. \textbf{23}, 117
(1987).

\bibitem{8} S. L. Woronowicz, Commun. Math. Phys. \textbf{111}, 613 (1987).

\bibitem{9} N. Seiberg and E. Witten, JHEP \textbf{9909}, 032 (1999).

\bibitem{10a} O. Bertolami and L. Guisado, JHEP \textbf{0312}, 013 (2003).

\bibitem{10} M. R. Douglas and N. A. Nekrasov, Rev. Mod. Phys. \textbf{73},
977 (2001).

\bibitem{11} M. Chaichian, M. M. Sheikh-Jabbari, and A. Tureanu, Phys. Rev.
Lett. \textbf{86}, 2716 (2001).

\bibitem{12} M. Chaichian, A. Demichev, P. Presnajder, M. M. Sheikh-Jabbari,
and A. Tureanu, Phys. Lett. \textbf{B 527}, 149 (2002).

\bibitem{13} R. J. Szabo, Physics Reports \textbf{378}, 207 (2003).

\bibitem{13a} S. Sivasubramanian, Y. N. Srivastava, G. Vitiello, and A.
Widom, Phys. Lett. \textbf{A 311}, 97 (2003).

\bibitem{14} M. Chaichian, M. M. Sheikh-Jabbari, and A. Tureanu, Eur. Phys.
J. \textbf{C 36}, 251 (2004).

\bibitem{14a} A. Kokado, T. Okamura and T. Saito, Phys. Rev. \textbf{D 69},
125007 (2004).

\bibitem{15} O. Bertolami, J. G. Rosa, C. M. L. de Arag$\check{\mathrm{a}}$%
o, P. Castorina, and D. Zappal$\grave{\mathrm{a}}$, Phys. Rev. \textbf{D 72}%
, 025010 (2005).

\bibitem{15n} K. Li, J. Wang, and C. Chen, Modern Physics Letters \textbf{A
20}, 2165 (2005).

\bibitem{15a} O. Bertolami, J. G. Rosa, C. M. L. de Arag$\check{\mathrm{a}}$%
o, P. Castorina, and D. Zappal$\grave{\mathrm{a}}$, Mod. Phys. Lett. \textbf{%
A 21}, 795 (2006).

\bibitem{16} N. Khosravi, H. R. Sepangi, and M. M. Sheikh-Jabbari, Phys.
Lett. \textbf{B 647}, 219 (2007).

\bibitem{17} M. Chaichian, A. Tureanu, and G. Zet, Phys. Lett. \textbf{B 660}%
, 573 (2008).

\bibitem{17a} C. Bastos, O. Bertolami, N. C. Dias, and J. N. Prata, Journal
of Mathematical Physics \textbf{49}, 072101 (2008).

\bibitem{18} M. M. Sheikh-Jabbari and A. Tureanu, Phys. Rev. Lett. \textbf{%
101}, 261601 (2008).

\bibitem{19} C. Bastos and O. Bertolami, Phys. Lett. \textbf{A 372}, 5556
(2008).

\bibitem{20} A. Alves and O. Bertolami, Phys. Rev. \textbf{D 82}, 047501
(2010).

\bibitem{21} C. Bastos, O. Bertolami, N. Costa Dias, and J. Nuno Prata,
Phys. Rev. \textbf{D 86}, 105030 (2012).

\bibitem{22} C. Bastos, O. Bertolami, N. Dias, and J. Prata, Int. J. Mod.
Phys. \textbf{A 28} 1350064 (2013).

\bibitem{23} A. E. Bernardini and O. Bertolami, Phys. Rev. \textbf{A 88},
012101 (2013).

\bibitem{89} S.-D. Liang, H. Li, and G.-Y. Huang, Phys. Rev. {\bf A 90}, 010102 (2014).

\bibitem{24} O. Bertolami and P. Leal, Phys. Lett. \textbf{B 750}, 6 (2015).

\bibitem{25} R. Bufalo and A. Tureanu, Phys. Rev. \textbf{D 92}, 065017
(2015).

\bibitem{26} C. Bastos, A. E. Bernardini, O. Bertolami, N. Costa Dias, and
J. Nuno Prata, Phys. Rev. \textbf{D 93}, 104055 (2016).

\bibitem{26a} A. A. Deriglazov and A. M. Pupasov-Maksimov, Phys. Lett. {\bf B 761},  207 (2016).

\bibitem{26b} A. A. Deriglazov and W. Guzman Ramirez, Advances in Mathematical Physics {\bf 2017}, 7397159 (2017).

\bibitem{26c} Kh. P. Gnatenko and V. M.  Tkachuk, Phys. Lett. {\bf A 381}, 2463 (2017).

\bibitem{26d} Kh. P.  Gnatenko, Europhysics Letters {\bf 123}, 50002 (2018).

\bibitem{26e} Kh. P.  Gnatenko, M. I. Samar, and V. M. Tkachuk, Phys. Rev. {\bf A 99}, 012114 (2019).

\bibitem{26f} Kh. P. Gnatenko,  Phys. Rev. {\bf D 99}, 026009 (2019).

\bibitem{27} S. Bellucci, A. Nersessian, and C. Sochichiu, Phys. Lett.
\textbf{B 522}, 345 (2001).

\bibitem{28} H. Falomir, J. Gamboa, J. Lopez-Sarrion, F. Mendez, and P. A.
G. Pisani, Phys. Lett. \textbf{B 680}, 384 (2009).

\bibitem{29} M. Gomes, V. G. Kupriyanov, and A. J. da Silva, Phys. Rev.
\textbf{D 81}, 085024 (2010).

\bibitem{30} A. Das, H. Falomir, J. Gamboa, F. Mendez, and M. Nieto, Phys.
Rev. \textbf{D 84}, 045002 (2011).

\bibitem{31} H. Falomir, J. Gamboa, M. Loewe, F. Mendez, and J. C. Rojas,
Phys. Rev. \textbf{D 85}, 025009 (2012).

\bibitem{32} A. F. Ferrari, M. Gomes, V. G. Kupriyanov, and C. A. Stechhahn,
Phys. Lett. \textbf{B 718}, 1475 (2013).

\bibitem{33} M. Gomes and V. G. Kupriyanov, Phys. Rev. \textbf{D 79} 125011
(2009).

\bibitem{34} M. Gomes, V. G. Kupriyanov, and A. J. da Silva, J. Phys. A:
Math. Theor. \textbf{43}, 285301 (2010).

\bibitem{35} A. Fring, L. Gouba, and F. G. Scholtz, J. Phys. \textbf{A 43},
345401 (2010).

\bibitem{36} V. G. Kupriyanov, J. Phys. A: Math. Theor. \textbf{46}, 245303
(2013).

\bibitem{37} V. G. Kupriyanov, J. Math. Phys. \textbf{54}, 112105 (2013).

\bibitem{37a} V. G. Kupriyanov, Phys. Lett. \textbf{B 732}, 385 (2014).

\bibitem{37b} V. G. Kupriyanov, Fortschr. Phys. \textbf{69}, 881 (2014).

\bibitem{38} S. A. Alavi and S. Abbaspour, J. Phys. A: Math. Theor. \textbf{%
47}, 045303 (2014).

\bibitem{39} S. A. Alavi and N. Rezaei, Pramana - J. Phys. \textbf{88}, 77
(2017).

\bibitem{40} S. A. Alavi and M. Amiri Nasab, Gen. Relativ. Gravit. \textbf{49%
}, 5 (2017).

\bibitem{40a} S. Dey and A. Fring, Phys. Rev. \textbf{D 90}, 084005 (2014).

\bibitem{41} L. Ts. Adzhemyan, T. L. Kim, M. V. Kompaniets, and V. K.
Sazonov, Nanosystems: physics, chemistry, mathematics \textbf{6}, 461 (2015).

\bibitem{End1} W. Pauli, Z. Phys. \textbf{43}, 601 (1927).

\bibitem{End2} J. Form\'{a}nek, R. J. Lombard, and J. Mare\^{s}, Czech. J.
Phys. \textbf{54}, 289 (2004).

\bibitem{End3} R. J. Lombard,J. Mare\^{s} and C. Volpe, J. Phys. G: Nucl.
Part. Phys. \textbf{34}, 1879 (2007).

\bibitem{End3a} M. De Sanctis and P. Quintero, Eur. Phys. J. A \textbf{39},
145 (2009).

\bibitem{End4} R. Yekken and R. J. Lombard, J. Phys. A: Math. Theor. \textbf{%
43}, 125301 (2010).

\bibitem{End5} R. Yekken, M. Lassaut, and R. J. Lombard, Annals of Physics
\textbf{338}, 195 (2013).

\bibitem{End6} A. Schulze-Halberg and O. Yesiltas, Journal of Mathematical
Physics \textbf{59}, 113503 (2018).

\bibitem{Stegun} M. Abramowitz and I. A. Stegun, Handbook of Mathematical
Functions: with Formulas, Graphs, and Mathematical Tables, Dover
Publications, INC., New York, 1965

\bibitem{Elbert} A. Elbert, Journal of Computational and Applied Mathematics
\textbf{133}, 65 (2001).

\bibitem{clim} S. H. Patil and K. D. Sen, Phys. Lett. {\bf A 362}, 109  (2007).

\bibitem{fract1} K. B. Oldham and J. Spanier, The Fractional Calculus:
Theory and Applications of Differentiation and Integration to Arbitrary
Order, Academic Press, Inc., New York, Dover Book Publications, 2006

\bibitem{fract2} M. D. Ortigueira, Fractional Calculus for Scientists and
Engineers (Lecture Notes in Electrical Engineering), Springer, Dordrecht,
Heidelberg, London, New York, 2011

\bibitem{Hermann} R. Herrmann, Fractional Calculus: An Introduction for
Physicists, World Scientific Publishing Company, Singapore, 2014

\bibitem{fSch1} N. Laskin, Phys. Rev. \textbf{E 66}, 056108 (2002).

\bibitem{fSch2} M. Naber, J. Math. Phys. \textbf{45}, 3339 (2004).

\bibitem{fSch3} J. Dong and M. Xu, Journal of Mathematical Physics \textbf{48%
}, 072105 (2007).

\bibitem{fSch4} M. Jeng, S.-L.-Y. Xu, E. Hawkins, and J. M. Schwarz, Journal
of Mathematical Physics \textbf{51}, 062102 (2010).

\bibitem{fSch5} A. N. Hatzinikitas, J. Math. Phys. \textbf{51}, 123523
(2010).

\bibitem{fSch6} M. Cheng, Journal of Mathematical Physics \textbf{53},
043507 (2012).

\bibitem{fSch7} S. S. Bayin, Journal of Mathematical Physics \textbf{54},
012103 (2013).

\bibitem{fSch8} B. A. Stickler, Phys. Rev. \textbf{E 88}, 012120 (2013).

\bibitem{fSch9} T. Sandev, Trifce, I. Petreska, and E. K. Lenzi, Journal of
Mathematical Physics \textbf{55}, 092105 (2014).

\bibitem{fSch9a} Z. Xiao, W. Chaozhen, L. Yingming, and L. Maokang, Annals
of Physics \textbf{350}, 124 (2014).

\bibitem{fSch10} Y. Zhang, X. Liu, M. R. Beli$\acute{\mathrm{c}}$, W. Zhong,
Y. Zhang, and M. Xiao, Phys. Rev. Lett. \textbf{115}, 180403 (2015).

\bibitem{fSch11} S. S. Bayin, Journal of Mathematical Physics \textbf{57},
123501 (2016).

\bibitem{fn1} D. Zhang, Y. Zhang, Z. Zhang, N. Ahmed, Y. Zhang, F. Li, M. R.
Belic, and M. Xiao, Annalen der Physik \textbf{529}, 1700149 (2017).

\bibitem{fn2} S. Bhattarai, Journal of Differential Equations \textbf{263},
3197 (2017).

\bibitem{fn3} A. Majlesi, H. Roohani Ghehsareh, and A. Zaghian, The European
Physical Journal Plus \textbf{132}, 516 (2017).

\bibitem{fn4} J. Li, Journal of Mathematical Physics \textbf{58}, 102701
(2017).

\bibitem{fn5} U. Al. Khawaja1, M. Al-Refai, G. Shchedrin and L. D. Carr,
Journal of Physics A: Mathematical and Theoretical \textbf{51}, 235201
(2018).

\bibitem{fn6} L. Shena and X. Yao, Journal of Mathematical Physics \textbf{59%
}, 081501 (2018).

\bibitem{fn7} M. Chen, S. Zeng, D. Lu, W. Hu, and Q. Guo, Phys. Rev. \textbf{%
E 98}, 022211 (2018).

\bibitem{fn8} X. Zhang, B. Yang, C. Wei, M. Luo, Communications in Nonlinear
Science and Numerical Simulation \textbf{67}, 290 (2019).

\bibitem{Laskin} N. Laskin, Fractional Quantum Mechanics, World Scientific, Singapore, 2018

\bibitem{La1} N. Laskin, Phys. Lett. \textbf{A 268}, 298 (2000).

\bibitem{La2} N. Laskin, Phys. Rev. \textbf{E 62}, 3135 (2000).


\end{thebibliography}
\end{document}